\documentclass{pasj00}
\draft

\begin{document}
\SetRunningHead{S. Kato}{Excitation of Disk Oscillations in Deformed Disks}
\Received{2007/0/00}
\Accepted{2007/0/00}

\title{Resonant Excitation of Disk Oscillations in Deformed Disks II:  
        A Model of High Frequency QPOs}

\author{Shoji \textsc{Kato}}%
\affil{2-2-2 Shikanodai-Nishi, Ikoma-shi, Nara 630-0114}
\email{kato.shoji@gmail.com; kato@kusastro.kyoto-u.ac.jp}


%


\maketitle

\begin{abstract}
The amplification of disk oscillations resulting from nonlinear resonant couplings
between the oscillations and a disk deformation is examined.
The disk is geometrically thin and general relativistic with a non-rotating central
source.
A Lagrangian formulation is adopted.
The author examined the same problem a few years ago, but here we derive
a general stability criterion in a more perspective way.
Another distinct point from the previous work is that 
in addition to the case where the deformation is a warp,
the case where the deformation is a one-armed pattern symmetric with respect 
to the equatorial plane is considered.
The results obtained show that in addition to the previous results that
the inertial-acoustic mode and g-mode oscillations are amplified by
horizontal resonance in  warped disks, they also amplified by 
horizontal resonance in disks deformed by one-armed pattern symmetric with
respect to the equatorial plane.
If we consider local oscillations that are localized around boundaries of their
propagation region,  the resonance occurs at $4r_{\rm g}$, where $r_{\rm g}$ is the 
Schwarzschild radius. 
If nonlocal oscillations are considered, frequency ranges of oscillations where
oscillations are resonantly amplified are specified. 

\end{abstract}


\section{Introduction}

Since the discovery of high-frequency quasi-periodic oscillations (QPOs)
by RXTE satellite in neutron-star X-ray and black-hole X-ray
binaries, much progress has been made both in observations of the high-frequency
QPOs and in their theoretical modeling.
This is because the QPOs are regarded to be a very powerful
tool to explore the mass and spin of the central compact sources and also
to explore the physical states of the innermost relativistic region of accretion disks.

One of promising models of the high-frequency QPOs is that they are disk
oscillations resonantly excited in relativistic disks.
At an early stage of observational development of the high-frequency
QPOs, Abramowicz and Klu{\' z}niak (2001) and Klu{\' z}niak and 
Abramowicz (2001) already pointed out the importance of resonant processes
as the cause of the high-frequency QPOs.
Their model is that the high-frequency QPOs are the results of resonant 
couplings between the vertical and horizontal epicyclic oscillations in disks.
This idea is very fascinating, but there seems to have still uncertainty concerning 
their excitation processes.

Kato (2003, 2004), on the other hand, proposed another resonant model of the
high-frequency QPOs, where the QPOs are regarded to be disk oscillations resonantly 
excited on {\it deformed} disks.
In this model, a deformation of disks from an axially-symmetric equilibrium state is an essential 
ingredient for resonant excitation of disk oscillations.
He especially considered the case where the disk deformation is a warp.
An outline of the model is as follows.
A nonlinear coupling between a disk oscillation (hereafter we call it the original oscillation) and a warp brings about some disk oscillations (we call them
intermediate oscillations).
The intermediate oscillations make a resonant coupling with the unperturbed disk at 
some particular radius of the disk.
After this resonant coupling with the unperturbed disk, the intermediate oscillations
feedback to the original oscillation by nonlinearly coupling with the warp.
Since the  nonlinear feedback processes involve a resonance, the original
oscillation is amplified or dampened.
For this nonlinear coupling to occur, however, some conditions must be satisfied.
These conditions specify the place(s) where the resonance occurs and the frequencies of
resonant oscillations (Kato 2004).
Among such resonant oscillations, some ones are amplified and others are 
dampened (Kato 2004).
Application of these resonantly-excited oscillations to the 
high-frequency QPOs has been made by Kato and his collaborators, e.g.,
Kato and Fukue (2006) and Kato (2007).

Examination in which cases the oscillations 
are resonantly excited was made by Kato (2003) and (2004), but their analyses 
are very lengthy and mathematically not perspective.
The reason is that cylindrical coordinates are introduced at a early stage of formulation.
In spite of this, the results obtained are rather simple.
This suggests that a more perspective and general derivation of stability criterion is possible.
Really, we find that the stability criterion can be derived in a simpler and 
perspective way.
A purpose of this paper is to show this.
The results obtained confirm the stability criteria obtained by Kato (2004).

In addition, we show in this paper that a warp is not only one type of deformation 
that can resonantly amplify disk oscillations.
One-armed deformations of disks that are symmetric with respect to the equatorial plane 
are also one of possible candidates that can resonantly amplify 
disk oscillations.

In our previous work, we specifically consider the cases where 
oscillations are local and localized around the boundaries of their propagation
region.
In this paper we relax this approximation, and specify the frequency ranges 
(in other words, the radial ranges) where global oscillations are resonantly amplified. 

\section{Summary of Basic Nonlinear Hydrodynamic Equations}

Here, we summarize basic nonlinear hydrodynamic equations that
are necessary to examine nonlinear resonant processes and their
stability.
Details are given by Kato (2004).

We are interested in general relativistic disks, since in the resonant
instability processes that we consider here the general relativity is essential.
The essence of the effects of the general relativity, however, can be taken into account within 
the framework of a pseudo-Newtonian formulation using the 
gravitational potential introduced by Paczy{\' n}ski and Wiita (1980),
provided that the central object has no rotation.

The formulation presented here are quite general until some 
approximations are introduced later.
The unperturbed state of the systems considered here
is a steady equilibrium state.
In Lagrangian formulation, the hydrodynamic equation describing
perturbations over the state can be written as (Lynden-Bell and Ostriker 1967)
\begin{equation}
    {D_0^2\mbox{\boldmath $\xi$}\over Dt^2}
    =\delta\biggr(-\nabla\psi-{1\over\rho}\nabla p\biggr),
\label{2.1}
\end{equation}
where $\mbox{\boldmath $\xi$}$ is a displacement vector associated 
with perturbations, and $D_0/Dt$ is the time derivative along an 
unperturbed flow, $\mbox{\boldmath $u$}_0$, and is related to the
Eulerian derivative, $\partial/\partial t$, by
\begin{equation}
   {D_0\over Dt}={\partial \over\partial t}+\mbox{\boldmath $u$}_0\cdot\nabla.
\label{2.2}
\end{equation}
In equation (\ref{2.1}), $\delta$ represents the Lagrangian variation
of the quantities in the subsequent parentheses, and $\psi$ is
the gravitational potential.
Other notations in equation (\ref{2.1}) have their usual meanings.

 The basic equation (\ref{2.1}) is valid even when the displacement 
$\mbox{\boldmath $\xi$}$ has a finite amplitude.
The purpose here is to explicitly write down the right-hand side of
equation (\ref{2.1}) up to the second-order quantities with respect to 
$\mbox{\boldmath $\xi$}$.
Now, we consider a system consisting of a central object and a disk 
surrounding the central object.
Perturbations are only on the disk and the self-gravity of the disk gas
is neglected.
That is, the gravitational potential, $\psi(\mbox{\boldmath $r$})$, comes
from the central object, and there is no Eulerian perturbation of
$\psi(\mbox{\boldmath $r$})$.\footnote{
The Lagrangian formulation by Lynden-Bell and Ostriker (1967) is
quite general, and the effects of self-gravity can be taken into account.
In this paper, however, we neglect the effects of self-gravity, since we are
interested in standard accretion disks.
}

To represent the Lagrangian variation of density, $\delta\rho$, by quantities 
up to the second-order ones with respect to $\mbox{\boldmath $\xi$}$,
we need the equation of continuity.
We assume here that the perturbations are adiabatic.
Then, the Lagrangian variation of pressure, $\delta p$, can be expressed
by quantities up to the second-order ones with respect to 
$\mbox{\boldmath $\xi$}$.
Then, the nonlinear hydrodynamic equation describing adiabatic, 
Nonself-gravitating perturbations is written as, after lengthy manipulation,
\begin{equation}
    \rho_0{\partial^2\mbox{\boldmath $\xi$}\over\partial t^2}
         +2\rho_0(\mbox{\boldmath $u$}_0\cdot\nabla)
                 {\partial\mbox{\boldmath $\xi$}\over\partial t}
      +\mbox{\boldmath $L$}(\mbox{\boldmath $\xi$})
      =\rho_0\mbox{\boldmath $C$}(\mbox{\boldmath $\xi$},
                                                       \mbox{\boldmath $\xi$}),
\label{2.3}
\end{equation}
where $\mbox{\boldmath $L$}(\mbox{\boldmath $\xi$})$ is a linear
operator with respect to $\mbox{\boldmath $\xi$}$ and is
(Lynden-Bell and Ostriker 1967)
\begin{eqnarray}
   \mbox{\boldmath $L$}(\mbox{\boldmath $\xi$})
     =\rho_0(\mbox{\boldmath $u$}_0\cdot\nabla)
                         (\mbox{\boldmath $u$}_0\cdot\nabla)\mbox{\boldmath $\xi$}
       +\rho_0(\mbox{\boldmath $\xi$}\cdot\nabla)(\nabla\psi_0)
       +\nabla\biggr[(1-\Gamma_1)p_0{\rm div}\mbox{\boldmath $\xi$}\biggr] 
                          \nonumber \\
       -p_0\nabla({\rm div}\mbox{\boldmath $\xi$})
       -\nabla[(\mbox{\boldmath $\xi$}\cdot\nabla)p_0]
       +(\mbox{\boldmath $\xi$}\cdot\nabla)(\nabla p_0),
\label{2.4}
\end{eqnarray}
and $\rho_0(\mbox{\boldmath $r$})$ and $p_0(\mbox{\boldmath $r$})$ are
the density and pressure in the unperturbed state, and $\Gamma_1$ is
the barotropic index specifying the linear part of the relation between 
Lagrangian variations $\delta p$ and $\delta \rho$, i.e.,
$(\delta p/p_0)_{\rm linear}=\Gamma_1(\delta\rho/\rho_0)_{\rm linear}$.
The right-hand side of wave equation (\ref{2.3}) represents the nonlinear terms, 
which consist of two parts, i.e., 
$\rho_0\mbox{\boldmath $C$}=\rho_0\mbox{\boldmath $C$}_\psi
     +\rho_0\mbox{\boldmath $C$}_{\rm p}$.
The former comes from $-\rho_0\delta(\nabla\psi)$ in equation (\ref{2.1})
and $\rho_0\mbox{\boldmath $C$}_p$ from $-\rho_0\delta(\nabla p/\rho)$.
Here and hereafter, for simplicity, we consider the case of $\Gamma_1=1$.
Then, we have (Kato 2004) (an expression for nonlinear terms in the general case 
of $\Gamma_1\not= 1$  is given in appendix)
\begin{equation}
     \rho_0\mbox{\boldmath $C$}_\psi=-{1\over 2}\rho_0\xi_i\xi_j
        {\partial ^2\over \partial r_i \partial r_j}(\nabla\psi_0)
\label{2.5}
\end{equation}
and
\begin{equation}
     \rho_0\mbox{\boldmath $C$}_{\rm p} =-{\partial\over\partial r_i}
           \biggr(p_0{\partial \xi_i\over \partial r_j}\nabla\xi_j\biggr).
\label{2.6}
\end{equation}

\section{Wave Energy and  Work Done on Oscillations}

We now assume that the disks are deformed from an axisymmetric 
steady state by some external or internal cause.
The Lagrangian displacement associated with the deformation is denoted by 
$\mbox{\boldmath $\xi$}^{\rm W}(\mbox{\boldmath $r$}, t)$.
The deformation may be time-periodic with frequency 
$\omega^{\rm W}$ as $\mbox{\boldmath $\xi$}^{\rm W}={\rm exp}
(i\omega^{\rm W}t)\hat{ \mbox{\boldmath $\xi$}}^{\rm W}
(\mbox{\boldmath $r$})$.
In order to avoid unnecessary complication, however, we take 
$\omega^{\rm W}=0$, hereafter.
The effects of non-zero $\omega^{\rm W}$ can easily be taken into account in the
final results.
As the deformation, a warp will be most probable, but it is not only one 
candidate of possible deformation.
A plane-symmetric one-armed spiral deformation is one of another possible 
candidates that excite disk oscillations, as discussed later.

Our purpose here is to examine how the behavior of disk oscillations 
are affected by the disk deformation.
As shown below, some of oscillation modes are resonantly excited
on deformed disks through nonlinear coupling with disk deformation.
The nonlinear coupling processes are schematically shown in figure 1
of Kato (2004).

Here we consider a disk oscillation mode.
The displacement vector associated with the oscillation is denoted by 
$\mbox{\boldmath $\xi$}$ with frequency $\omega$, i.e., 
$\mbox{\boldmath $\xi$}={\rm exp}(i\omega t)
\hat {\mbox{\boldmath $\xi$}}(\mbox{\boldmath $r$})$, in the limit of 
no deformation of disks.
The first step of the nonlinear interaction between the disk oscillation
characterized by $(\omega, \hat {\mbox{\boldmath $\xi$}})$
and the deformation characterized by $(0,
\hat{\mbox{\boldmath $\xi$}}^{\rm W})$ introduces two kinds 
of intermediate oscillations.
Let us denote the displacement vector associated with these intermediate
oscillations by $\mbox{\boldmath $\xi$}_\pm^{\rm int}
[\equiv{\rm exp}( i \omega t)
\hat {\mbox{\boldmath $\xi$}}_\pm^{\rm int}]$.
In other words, we characterize the two intermediate oscillations by
$(\omega, \hat{\mbox{\boldmath $\xi$}}_+^{\rm int})$ and
$(\omega, \hat{\mbox{\boldmath $\xi$}}_-^{\rm int})$, where
$\hat{\mbox{\boldmath $\xi$}}_+^{\rm int}$ represents the intermediate 
oscillations resulting from the coupling between 
$\hat {\mbox{\boldmath $\xi$}}$ and 
$\hat {\mbox{\boldmath $\xi$}}^{\rm W}$, while 
$\hat {\mbox{\boldmath $\xi$}}_-^{\rm int}$ does those resulting from
the coupling between $\hat {\mbox{\boldmath $\xi$}}$ and
$\hat {\mbox{\boldmath $\xi$}}^{{\rm W}*}$, where the asterisk 
represents the complex conjugate.
That is, $\hat {\mbox{\boldmath $\xi$}}_+^{\rm int}$ and
$\hat {\mbox{\boldmath $\xi$}}_-^{\rm int}$ are described, respectively, by
\begin{equation}
   -\omega^2\rho_0\hat{\mbox{\boldmath $\xi$}}_+^{\rm int}
  +2i\omega\rho_0(\mbox{\boldmath $u$}_0\cdot\nabla)
                      \hat{ \mbox{\boldmath $\xi$}}_+^{\rm int}
  +\mbox{\boldmath $L$}(\hat{\mbox{\boldmath $\xi$}}_+^{\rm int})
   ={1\over 2}[\rho_0\mbox{\boldmath $C$}(\hat{\mbox{\boldmath $\xi$}},
                                  \hat{\mbox{\boldmath $\xi$}}^{\rm W})
                     +\rho_0\mbox{\boldmath $C$}
                                  (\hat{\mbox{\boldmath $\xi$}}^{\rm W},  
                                  \hat{\mbox{\boldmath $\xi$}})].
\label{2.7}
\end{equation}
\begin{equation}
   -\omega^2\rho_0\hat{\mbox{\boldmath $\xi$}}_-^{\rm int}
  +2i\omega\rho_0(\mbox{\boldmath $u$}_0\cdot\nabla)
                      \hat{ \mbox{\boldmath $\xi$}}_-^{\rm int}
  +\mbox{\boldmath $L$}(\hat{\mbox{\boldmath $\xi$}}_-^{\rm int})
        = {1\over 2}[\rho_0\mbox{\boldmath $C$}
                           (\hat{\mbox{\boldmath $\xi$}},
                                  \hat{\mbox{\boldmath $\xi$}}^{{\rm W}*})
                     +\rho_0\mbox{\boldmath $C$}
                                  (\hat{\mbox{\boldmath $\xi$}}^{{\rm W}*},  
                                   \hat{\mbox{\boldmath $\xi$}})].
\label{2.8}
\end{equation}

Next, the second stage of the nonlinear coupling is considered, which is a 
feedback process returning to the original oscillation, $\mbox{\boldmath $\xi$}$,
by $\mbox{\boldmath $\xi$}_+^{\rm int}$ (or
$\mbox{\boldmath $\xi$}_-^{\rm int})$ interacting with
$\mbox{\boldmath $\xi$}^{\rm W}$.
The feedback is described by
\begin{equation}
   -\omega^2\rho_0\hat{\mbox{\boldmath $\xi$}}
  +2i\omega\rho_0(\mbox{\boldmath $u$}_0\cdot\nabla)
                      \hat{ \mbox{\boldmath $\xi$}}
  +\mbox{\boldmath $L$}(\hat{\mbox{\boldmath $\xi$}})
   ={1\over 2}[\rho_0\mbox{\boldmath $C$}
                   (\hat{\mbox{\boldmath $\xi$}}_+^{\rm int},
                    \hat{\mbox{\boldmath $\xi$}}^{{\rm W}*})
                    +\rho_0\mbox{\boldmath $C$}
                    (\hat{\mbox{\boldmath $\xi$}}^{{\rm W}*},  
                     \hat{\mbox{\boldmath $\xi$}}_+^{\rm int})],
\label{2.9}
\end{equation}
in the case where the nonlinear coupling between 
$\hat{\mbox{\boldmath $\xi$}}_+^{\rm int}$ and 
$\hat{\mbox{\boldmath $\xi$}}^{\rm W}$ feedbacks to the original oscillation,
$\mbox{\boldmath $\xi$}$.
On the other hand, in the case where $\mbox{\boldmath $\xi$}_-^{\rm int}$
and $\mbox{\boldmath $\xi$}^{\rm W}$ couples to feedback to the 
original oscillation, the equation corresponding to equation (\ref{2.9}) is 
\begin{equation}
       -\omega^2\rho_0\hat{\mbox{\boldmath $\xi$}}
           +2i\omega\rho_0(\mbox{\boldmath $u$}_0\cdot\nabla)
                      \hat{ \mbox{\boldmath $\xi$}}
           +\mbox{\boldmath $L$}(\hat{\mbox{\boldmath $\xi$}})
       = {1\over 2}[\rho_0\mbox{\boldmath $C$}
                   (\hat{\mbox{\boldmath $\xi$}}_-^{\rm int},
                    \hat{\mbox{\boldmath $\xi$}}^{{\rm W}})
                    +\rho_0\mbox{\boldmath $C$}
                    (\hat{\mbox{\boldmath $\xi$}}^{{\rm W}},  
                     \hat{\mbox{\boldmath $\xi$}}_-^{\rm int})].
\label{2.10}
\end{equation}

An important point to be noted here is that as a result of this feedback
process, the original disk oscillation is amplified or dampened, i.e.,
the frequency, $\omega$, can be no longer real, since in the feedback 
a resonant process is involved as shown later.
How much is the imaginary part of $\omega$?
This can be examined from equations (\ref{2.9}) and (\ref{2.10}) 
by using the fact that the operators $i\rho_0(\mbox{\boldmath{$u$}}_0\cdot\nabla)$ and  
$\mbox{\boldmath $L$}$ are Hermitian (Lynden-Bell and Ostriker 1967).

Let us now multifly equation (\ref{2.9}) by $\hat{\mbox{\boldmath $\xi$}}^*$
and integrate over the whole volume, and consider the imaginary part of
the resulting equation.
The volume integration of $\hat{\mbox{\boldmath $\xi$}}^*\cdot
\mbox{\boldmath $L$}(\hat {\mbox{\boldmath $\xi$}})$ does not have
any imaginary part, since the operator $\mbox{\boldmath $L$}$ is
Hermitian.
Now we write $\omega=\omega_0+i\omega_i$, where $\omega_i$ is the
imaginary part of $\omega$.
Then, from equation (\ref{2.9}) we have
\begin{equation}
     -2\omega_{\rm i}\int\rho_0\mbox{\boldmath $\xi$}^*
            [\omega_0-i(\mbox{\boldmath $u$}_0\cdot\nabla)]\mbox{\boldmath $\xi$}dV
      =\Im {1\over 2}\int\rho_0\hat{\mbox{\boldmath $\xi$}}^*
            [\mbox{\boldmath $C$}(\hat{\mbox{\boldmath $\xi$}}_+^{\rm int},
                   \hat{\mbox{\boldmath $\xi$}}^{{\rm W}*})
                +\mbox{\boldmath $C$}(\hat{\mbox{\boldmath $\xi$}}^{{\rm W}*},
                   \hat{\mbox{\boldmath $\xi$}}_+^{\rm int})]dV.
\label{2.11}
\end{equation}
This equation has a clear physical meaning.
The integral on the left-hand side is related to the wave energy.
Really, the wave energy is expressed as (Kato 2001)
\begin{equation}
       E={1\over 2}\omega_0\int\rho_0\hat{\mbox{\boldmath $\xi$}}^*
               [\omega_0-i(\mbox{\boldmath $u$}\cdot\nabla)]
                        \hat{\mbox{\boldmath $\xi$}}dV.
\label{2.12}
\end{equation}
The right-hand side of equation (\ref{2.11}) is thus 
$-4(\omega_{\rm i}/\omega_0)E$.
Hereafter, the subscript 0 to $\omega$ is neglected for simplicity.
In the case of oscillations that have $m$-arms in the azimuthal direction
in geometrically thin disks, the wave energy $E$ can be expressed 
in terms of $\mbox{\boldmath $\xi$}$ as (Kato 2001)
\begin{equation}
    E={1\over 2}\int \omega(\omega-m\Omega)\rho_0(\vert\hat{\xi}_r\vert^2
              +\vert\hat{\xi}_z\vert^2)dV,
\label{2.13}
\end{equation}
where the cylindrical coordinates ($r$, $\varphi$, $z$) whose origin is at the
disk center and the $z$-axis is the axis of disk rotation have been introduced.

Next, the right-hand side of equation (\ref{2.11}) is considered.
This is related to the work done on oscillations.
Let us assume that an external force $\mbox{\boldmath $f$}$ 
on the oscillations acts per unit volume.
Then, the work done by the force on the oscillations per unit time, $W$,
is
\begin{equation}
       W=\langle \int\rho_0{\partial \mbox{\boldmath $\xi$}\over\partial t}\cdot
                   \mbox{\boldmath $f$}dV\rangle,
\label{2.14}
\end{equation}
where $\langle$ $\rangle$ represents the time average.
If we write $\mbox{\boldmath $\xi$}=\Re [{\rm exp}(i\omega t)
    \hat{\mbox{\boldmath $\xi$}}]$ and 
$\mbox{\boldmath $f$}=\Re [{\rm exp}(i\omega t)
    \hat{\mbox{\boldmath $f$}}]$, we find that $W$ can be expressed as
\begin{equation}
    W ={\omega\over 2} \Im \int\rho_0\hat{\mbox{\boldmath $\xi$}}^*
               \cdot\hat{\mbox{\boldmath $f$}}dV.
\label{2.15}
\end{equation}
This means that in the present case, i.e., equation (\ref{2.9}), the work done on the 
original oscillations by the nonlinear coupling terms is 
\begin{equation}
    W_+={\omega\over 2}\Im\int {1\over 2}\rho_0\hat{\mbox{\boldmath $\xi$}}^*
       [\mbox{\boldmath $C$}(\hat{\mbox{\boldmath $\xi$}}_+^{\rm int},
                        \hat{\mbox{\boldmath $\xi$}}^{{\rm W}*})
     + \mbox{\boldmath $C$}(\hat{\mbox{\boldmath $\xi$}}^{{\rm W}*},
                        \hat{\mbox{\boldmath $\xi$}}_+^{\rm int})] dV,
\label{2.16}
\end{equation}
where the subscript + is attached to $W$ in order to emphasize that 
the feedback occurs through $\hat{\mbox{\boldmath{$\xi$}}}_+^{\rm int}$.
Hence, equation (\ref{2.11}) can be expressed as
\begin{equation}
      -\omega_{{\rm i},+} ={W_+\over 2E},
\label{2.17}
\end{equation}
which is an expected result.

In the case of the coupling through 
$\hat{\mbox{\boldmath $\xi$}}_-^{\rm int}$, the work done the original 
oscillations is written as
\begin{equation}
    W_-={\omega\over 2} \Im\int {1\over 2}\rho_0\hat{\mbox{\boldmath $\xi$}}^*
       [\mbox{\boldmath $C$}(\hat{\mbox{\boldmath $\xi$}}_-^{\rm int},
                        \hat{\mbox{\boldmath $\xi$}}^{{\rm W}})
     + \mbox{\boldmath $C$}(\hat{\mbox{\boldmath $\xi$}}^{{\rm W}},
                        \hat{\mbox{\boldmath $\xi$}}_-^{\rm int})] dV,
\label{2.18}
\end{equation}
and 
\begin{equation}
      -\omega_{{\rm i},-} ={W_-\over 2E}.
\label{2.19}
\end{equation}
The quantitities $\omega_{\rm i}$'s in equations (\ref{2.17}) and (\ref{2.19})
are different by difference of coupling path.
In order to distinguish them, the subscript $+$ or $-$ is attached to 
$\omega_{\rm i}$.
 
The purpose in the following sections is thus to express $W_\pm$ more
explicitly and to evaluate the sigh of $W_\pm$.
It is noted that there are two cases where the wave energy is positive 
and negative.
If the waves exist dominantly inside the corotation radius (i.e., the radius of
$\omega=m\Omega$), the wave energy $E$ is negative [see equation (\ref{2.13})]
and a negative
work ($W<0$) leads to instability ($-\omega_{\rm i}>0$), while
the waves exist mainly outside the corotation radius, we have
$E>0$, and $W>0$ leads to instability.

The coupling term $\mbox{\boldmath $C$}$ consists of two
terms of  $\mbox{\boldmath $C$}_\psi$ and
$\mbox{\boldmath $C$}_{\rm p}$,  given by equation (\ref{2.5})
and (\ref{2.6}), respectively.
From equaion (\ref{2.5}) we easily find that
\begin{equation}
    \int\rho_0\hat{\mbox{\boldmath $\xi$}}^*\mbox{\boldmath $C$}_\psi
         (\hat{\mbox{\boldmath $\xi$}}_+^{\rm int}, 
          \hat{\mbox{\boldmath $\xi$}}^{{\rm W}*})dV
    =\int\rho_0\hat{\mbox{\boldmath $\xi$}}_+^{\rm int}
                              \mbox{\boldmath $C$}_\psi
         (\hat{\mbox{\boldmath $\xi$}}^*, 
          \hat{\mbox{\boldmath $\xi$}}^{{\rm W}*})dV.
\label{3.1}
\end{equation}
In the case of the integration of $\rho_0\hat{\mbox{\boldmath $\xi$}}^*
\mbox{\boldmath $C$}_{\rm p}$ over the volume, we perform two 
times the integration by part.
Then, we have
\begin{equation}
    \int\rho_0\hat{\mbox{\boldmath $\xi$}}^*\mbox{\boldmath $C$}_{\rm p}
         (\hat{\mbox{\boldmath $\xi$}}_+^{\rm int}, 
          \hat{\mbox{\boldmath $\xi$}}^{{\rm W}*})dV
    =\int\rho_0\hat{\mbox{\boldmath $\xi$}}_+^{\rm int}
                              \mbox{\boldmath $C$}_{\rm p}
         (\hat{\mbox{\boldmath $\xi$}}^*, 
          \hat{\mbox{\boldmath $\xi$}}^{{\rm W}*})dV.
\label{3.2}
\end{equation}
Hence, $W_+$ given by equation (\ref{2.16}) is reduced to
\begin{equation}
   W_+={\omega\over 2}\Im \int{1\over 2}\rho_0\hat{\mbox{\boldmath $\xi$}}_+^{\rm int}
         [\mbox{\boldmath $C$}(\hat{\mbox{\boldmath $\xi$}}^*,
                  \hat{\mbox{\boldmath $\xi$}}^{{\rm W}*})
        +\mbox{\boldmath $C$}(\hat{\mbox{\boldmath $\xi$}}^{{\rm W}*},
                  \hat{\mbox{\boldmath $\xi$}}^*)]dV.
\label{3.3}
\end{equation}
It is very important to note here that $\hat{\mbox{\boldmath $\xi$}}^*$,
$\hat{\mbox{\boldmath $\xi$}}^{\rm int}$, and
$\hat{\mbox{\boldmath $\xi$}}^{\rm W}$ in integrations (\ref{3.1}) and (\ref{3.2})
are commnesurable each other as one of their examples is shown above.
This nature is not limited in the case of $\Gamma_1=1$.
This holds generally even in the case of $\Gamma_1\not= 1$, as shown in
appendix.
The terms in the brackets of equation (\ref{3.3}) are just the complex
conjugate of the right-hand side of equation (\ref{2.7}), which makes the 
calculation of $W_+$ easy, as shown later.

In the case of the feedback through $\hat{\mbox{\boldmath {$\xi$}}}_-^{\rm int}$,
we similarly have, from equation (\ref{2.18}),  
\begin{equation}
     W_-={\omega\over 2}\Im \int{1\over 2}\rho_0\hat{\mbox{\boldmath $\xi$}}_-^{\rm int}
         [\mbox{\boldmath $C$}(\hat{\mbox{\boldmath $\xi$}}^*,
                  \hat{\mbox{\boldmath $\xi$}}^{{\rm W}})
        +\mbox{\boldmath $C$}(\hat{\mbox{\boldmath $\xi$}}^{{\rm W}},
                  \hat{\mbox{\boldmath $\xi$}}^*)]dV.
\label{3.4}
\end{equation}
The terms in the brackets are just the complex conjugate of the right-hand side of 
equation (\ref{2.8}), which makes the calculation of $W_{\rm -}$ easy,
as in the case of $W_+$.

\section{Dispersion Relation}

Hereafter we restrict our attention to oscillations in geometrically thin disks.
The steady unperturbed disks are assumed to be axially-symmetric and
to have no motion except for rotation.
Here, cylindrical coordinates ($r$, $\varphi$, $z$) are employed, in which 
the $z$-axis is perpendicular to the disk plane and the origin of the
coordinates is at the disk center.
The rotation is then described as $\mbox{\boldmath $u$}_0=
(0,r\Omega(r),0)$.
When we need numerical figures, we adopt the Keplerian form of $\Omega$,
i.e., $\Omega=\Omega_{\rm K}$, since geometrically thin disks are
considered.

The oscillations that we consider here are assumed to have $m$ arms in the
azimuthal direction, i.e., $\hat{\mbox{\boldmath $\xi$}}\propto
{\rm exp}(-im\varphi)$.
We further assume that the radial wavelength of the oscillations is moderately short,
and the local approximation in the radial direction can be adopted.
That is, $\hat{\mbox{\boldmath $\xi$}}\propto{\rm exp}(ikr)$ and
the radial variations of physical quantities in the unperturbed state are
neglected compared with the radial variation of $\hat{\mbox{\boldmath $\xi$}}$.
Concerning the vertical direction, however, we cannot adopt local approximations,
since the disk is geometrically thin.

For mathematical simplicity, we assume that the disks are vertically isothermal.
In the vertically isothermal disks 
the density $\rho_0(r,z)$ is stratified as (e.g., Kato et al. 1998)
\begin{equation}
         \rho_0(r,z)=\rho_{00}(r){\rm exp}\biggr[-{z^2\over 2H^2(r)}\biggr],
\label{4.4}
\end{equation}
where $\rho_{00}$ is the density on the equatorial plane, and $H$
is the half-thickness of the disk and  is related to the vertical epicyclic frequency,
$\Omega_\bot$, by
\begin{equation}
       \Omega_\bot^2H^2={p_0\over \rho_0}=c^2_{\rm s}(r). 
\label{4.5}
\end{equation}
The vertical epicyclic frequency, $\Omega_\bot$, is equal to the angular velocity
of the Keplerian rotation, $\Omega_{\rm K}$, in the case of the central object 
being non-rotating, and practically equal to $\Omega$, since the disk is assumed to be geometrically thin.
Hereafter, however, we use $\Omega_\bot$ without using $\Omega_{\rm K}$ or
$\Omega$, so that we can trace back the effects of $\Omega_\bot$ on the final
results.  
Furthermore, we assume that oscillations also occur isothermally.   
Then, we can neglect the term of $\nabla[(1-\Gamma_1)p_0{\rm div}
\mbox{\boldmath $\xi$}]$ in the operator $\mbox{\boldmath $L$}$
in equation (\ref{2.4}).
Under these approximations, we express the $r$-, $\varphi$-, and
$z$- components of the homogeneous parts of wave equation (\ref{2.3}).
Then, we have
\begin{equation}
     [-(\omega-m\Omega)^2+\kappa^2-4\Omega^2+k^2c_{\rm s}^2]\hat{\xi}_r
         -i2\Omega(\omega-m\Omega)\hat{\xi}_\varphi
        -ik\biggr(c_{\rm s}^2{\partial\hat{\xi}_z\over\partial z}
                  -z\Omega^2_\bot\hat{\xi}_z\biggr)=0,
\label{4.1}
\end{equation}
\begin{equation}
      -(\omega-m\Omega)^2\hat{\xi}_\varphi
              +i2\Omega(\omega-m\Omega)\hat{\xi}_r=0,
\label{4.2}
\end{equation}
\begin{equation}
     -(\omega-m\Omega)^2\hat{\xi}_z
                   -ikc_{\rm s}^2{\partial\hat{\xi}_r\over\partial z}
     -\biggr(c_{\rm s}^2{\partial^2\hat{\xi}_z\over\partial z^2}
               -\Omega_\bot^2z{\partial\hat{\xi}_z\over\partial z}
                            -\Omega_\bot^2\hat{\xi}_z\biggr)=0.
\label{4.3}
\end{equation}

These wave equations are differential equations, and it is troublesome to 
rigorously solve these equations.
In vertically isothermal disks, however, the wave equations can be approximately 
solved by separating $r$ and $z$ dependences of $\hat{\xi}$'s as,
for example, $\hat{\xi}_r(r,z)=f(r)g(z/H)$, where $f$ and $g$ are some functions
of $r$ and $z/H$, respectively (Okazaki et al. 1987).
Here, the functional form of $g(z/H)$ is determined by solving an
eigenvalue problem describing the vertical behavior of the oscillations, and
we have
\begin{equation}
             g(z)\propto {\cal H}_n\biggr({z\over H}\biggr), 
\label{4.6}
\end{equation}
where ${\cal H}_n$ is the Hermite polynomial of argument $z/H$, and
$n(=0,1,2,...)$ characterizes the number of node(s) of   oscillations
in the vertical direction.
More explicitly, we have
\begin{equation}
      \hat{\xi}_r(\mbox{\boldmath $r$})=\breve{\xi}_{r,n}(r,\varphi){\cal H}_n(z/H)                        
\label{4.7}
\end{equation}
\begin{equation}
      \hat{\xi}_\varphi(\mbox{\boldmath $r$}) 
                               = \breve{\xi}_{\varphi,n}(r,\varphi){\cal H}_n(z/H),
\label{4.8}
\end{equation}
\begin{equation}
      \hat{\xi}_z(\mbox{\boldmath $r$})=\breve{\xi}_{z, n}(r,\varphi){\cal H}_{n-1}(z/H).
\label{4.9}
\end{equation}
It is noted that the number of node(s) of $\hat{\xi}_z$ in the vertical
direction is smaller than those of $\hat{\xi}_r$ and $\hat{\xi}_\varphi$
by one, as shown in equation (\ref{4.9}).
However, the subscript $n$ (not $n-1$) is attached to $\breve{\xi}_z$ 
as $\breve{\xi}_{z,n}$ in order to emphasize that $\breve{\xi}_{r,n}$, 
$\breve{\xi}_{\varphi,n}$,  and $\breve{\xi}_{z, n}$ are a set of
solutions.
The fact that the $z$-dependences of $\hat{\xi}_r$, $\hat{\xi}_\varphi$,
and $\hat{\xi}_z$ given by equations (\ref{4.7}) -- (\ref{4.9}) are really
solutions of the set of equations (\ref{4.1}) -- (\ref{4.3}) is found by
substituting equations (\ref{4.7}) -- (\ref{4.9}) into equations
(\ref{4.1}) -- (\ref{4.3}) and conforming that the right-hand side of 
equations (\ref{4.1}) -- (\ref{4.3}) can be packed into terms proportional
to ${\cal H}_n$, ${\cal H}_n$, and ${\cal H}_{n-1}$, respectively.
The results are (Kato 2004)
\begin{equation}
      [-(\omega-m\Omega)^2+\kappa^2-4\Omega^2+k^2c_{\rm s}^2]\breve{\xi}_{r,n}
        -i2\Omega(\omega-m\Omega)\breve{\xi}_{\varphi,n}
        +i(kH)\Omega_\bot^2\breve{\xi}_{z,n}=0,
\label{4.10}
\end{equation}
\begin{equation}
      -(\omega-m\Omega)^2\breve{\xi}_{\varphi,n}
              +i2\Omega(\omega-m\Omega)\breve{\xi}_{r,n}=0,
\label{4.11}
\end{equation}
\begin{equation}
    [-(\omega-m\Omega)^2+n\Omega_\bot^2]\breve{\xi}_{z,n}
          -in(kH)\Omega_\bot^2\breve{\xi}_{r,n}=0.
\label{4.12}
\end{equation}

The solvability condition of these homogeneous equations is
\begin{equation}
        D(\omega,m,n)\equiv[(\omega-m\Omega)^2-\kappa^2]
                    [(\omega-m\Omega)^2-n\Omega_\bot^2]
                 -k^2c_{\rm s}^2(\omega-m\Omega)^2=0.
\label{4.13}
\end{equation}
This is the well-known local dispersion relation of oscillations characterized by
$\omega$, $m$, and $n$.

\section{Expressions for Work Done on Oscillations and Stability Conditions}

Our purpose here is to evaluate $W_\pm$ given by equations (\ref{3.3}) and
(\ref{3.4}).
As disk deformation, we consider two cases:
\begin{itemize}
\item
   case (i): $m^{\rm W}=1$ and $n^{\rm W}=1$.
\item
   case (ii):  $m^{\rm W}=1$ and $n^{\rm W}=0$.
\end{itemize}
The former is a warp and the latter is a one-armed pattern plane-symmetric 
with respect to the equatorial plane.  

In both cases, for simplicity, the radial wavelength of the deformation is
assumed to be rather long compared with the radial wavelength of the
disk oscillations.
That is, the radial variation of $\hat{\mbox{\boldmath $\xi$}}^{\rm W}$ is  
neglected compared with that of $\hat{\mbox{\boldmath $\xi$}}$.
For convenience, the part of the $z$-dependence of 
$\hat{\mbox{\boldmath $\xi$}}^{\rm W}$
is separated from other parts as
\begin{equation}
     \hat{\xi}_r^{\rm W}(\mbox{\boldmath $r$})
                        =\breve{\xi}_r(r,\varphi){\cal H}_{n^{\rm W}}(z/H),
\label{5.1}
\end{equation}
\begin{equation}
     \hat{\xi}_\varphi^{\rm W}(\mbox{\boldmath $r$})
                       =\breve{\xi}_\varphi(r,\varphi) {\cal H}_{n^{\rm W}}(z/H),
\label{5.2}
\end{equation}
\begin{equation}
     \hat{\xi}_z^{\rm W}(\mbox{\boldmath $r$})
                      =\breve{\xi}_z(r,\varphi) {\cal H}_{n^{\rm W}-1}(z/H),
\label{5.3}
\end{equation}
where $n^{\rm W}=1$ in case (i), and $n^{\rm W}=0$ and $\breve{\xi}_z$
is case (ii).

\subsection{Inhomogeneous Wave Equation for Intermediate Oscillations}

Nonlinear coupling between  $\mbox{\boldmath $\xi$}$
and $\mbox{\boldmath $\xi$}^{\rm W}$  introduces intermediate oscillations,
represented by $\mbox{\boldmath $\xi$}^{\rm int}= {\rm exp}(i\omega t)
\hat{\mbox{\boldmath $\xi$}}^{\rm int}$.
Related to the fact that ${\cal H}_1{\cal H}_n={\cal H}_{n+1}+n{\cal H}_{n-1}$, there are two
major normal modes of intermediate oscillations induced by the coupling between 
oscillations and a warp.
They are those characterized by ${\cal H}_{n+1}(z/H)$
and those characterized by ${\cal H}_{n-1}(z/H)$.\footnote{
In case (ii), the $z$-dependence of the intermediate oscillations is ${\cal H}_n(z/H)$ alone.
}
Concerning the $\varphi$-dependence of $\hat{\mbox{\boldmath $\xi$}}^{\rm int}$,
there are also two cases where $\hat{\mbox{\boldmath $\xi$}}^{\rm int}$ 
is proportional to ${\rm exp}[-i(m+1)\varphi]$
and to ${\rm exp}[-(m-1)\varphi]$.
That is, there are four modes of $\hat{\mbox{\boldmath $\xi$}}^{\rm int}$
by differences of the $\varphi$- and 
$z$-dependences in the case of a warp.
In order to treat these four modes separately, 
$\hat{\mbox{\boldmath $\xi$}}^{\rm int}$ is written as
\begin{equation}
     \hat{\xi}_r^{\rm int}(\mbox{\boldmath $r$})
          =\sum \breve{\xi}^{\rm int}_{r,\pm,\tilde{n}}(r,\varphi){\cal H}_{\tilde{n}}(z/H),
\label{5.4}
\end{equation}
\begin{equation}
     \hat{\xi}_\varphi^{\rm int}(\mbox{\boldmath $r$})
         =\sum \breve{\xi}^{\rm int}_{\varphi,\pm,\tilde{n}}(r,\varphi) 
                       {\cal H}_{\tilde{n}}(z/H),
\label{5.5}
\end{equation}
\begin{equation}
     \hat{\xi}_z^{\rm int}(\mbox{\boldmath $r$})
            =\sum \breve{\xi}^{\rm int}_{z,\pm,\tilde{n}} (r,\varphi)
                    {\cal H}_{\tilde{n}-1}(z/H),
\label{5.6}
\end{equation}
where the subscript $\pm$ is attached to $\breve{\xi}^{\rm int}$ to represent the two 
cases where $\breve{\mbox{\boldmath $\xi$}}^{\rm int}$ is proportional to 
${\rm exp}[-i(m\pm 1)\varphi]$ (see also section 3), and the subscript 
$\tilde{n}$ represents 
$n+1$ or $n-1$ in case (i) and $\tilde{n}=n$ in case (ii).

Now we solve inhomogeneous wave equation (\ref{2.7}) of 
 $\hat{\mbox{\boldmath $\xi$}}^{\rm int}_+$.
All the terms on the right-hand side of equation (\ref{2.7}) have
the $\varphi$-dependence of ${\rm exp}[-i(m+1)\varphi]$.
Their $z$-dependences, however, are not always limited to 
${\cal H}_{\tilde{n}}(z/H)$ nor ${\cal H}_{\tilde{n}-1}(z/H)$, although the   
major terms are those with such $z$-dependences.
Here, we write the right-hand side of equation (\ref{2.7}) as
\begin{equation}
      {1\over 2}\rho_0[ \mbox{\boldmath $C$}( \hat{\mbox{\boldmath $\xi$}},
                     \hat{\mbox{\boldmath $\xi$}}^{\rm W})
              + \mbox{\boldmath $C$}( \hat{\mbox{\boldmath $\xi$}}^{\rm W},
                      \hat{\mbox{\boldmath $\xi$}})]_r
       =\rho_0\sum_{\tilde{n}} \breve{A}_{r,+,\tilde{n}}(r, \varphi)
                {\cal H}_{\tilde {n}}(z/H)+\cdot\cdot\cdot
\label{5.7}
\end{equation}
\begin{equation}
      {1\over 2}\rho_0[ \mbox{\boldmath $C$}( \hat{\mbox{\boldmath $\xi$}},
                     \hat{\mbox{\boldmath $\xi$}}^{\rm W})
              + \mbox{\boldmath $C$}( \hat{\mbox{\boldmath $\xi$}}^{\rm W},
                      \hat{\mbox{\boldmath $\xi$}})]_\varphi
       =\rho_0\sum_{\tilde{n}} \breve{A}_{\varphi,+,\tilde{n}}(r,\varphi)
                {\cal H}_{\tilde {n}}(z/H)+\cdot\cdot\cdot
\label{5.7'}
\end{equation}
\begin{equation}
      {1\over 2}\rho_0[ \mbox{\boldmath $C$}( \hat{\mbox{\boldmath $\xi$}},
                     \hat{\mbox{\boldmath $\xi$}}^{\rm W})
              + \mbox{\boldmath $C$}( \hat{\mbox{\boldmath $\xi$}}^{\rm W},
                      \hat{\mbox{\boldmath $\xi$}})]_z
       =\rho_0\sum_{\tilde{n}} \breve{A}_{z,+,\tilde{n}}(r,\varphi)
                {\cal H}_{\tilde {n}-1}(z/H)+\cdot\cdot\cdot,
\label{5.7''}
\end{equation}
and concentrate our attention on the terms shown by $\breve{A}$'s.
Here, $+\cdot\cdot\cdot$ denotes other terms orthogonal both to 
${\cal H}_{\tilde{n}}$ and
${\cal H}_{\tilde{n}-1}$, and the subscript $+$ is added to $\breve{A}$'s 
in order to emphasize that they have the $\varphi$-dependence of 
${\rm exp}[-i(m+1)\varphi]$.

Under these preparations, we obtain from equation (\ref{2.7}) equations 
describing $\breve{\xi}_{r,+,\tilde{n}}$, $\breve{\xi}_{\varphi,+,\tilde{n}}$,
and $\breve{\xi}_{z,+,\tilde{n}}$ as [cf., equations (\ref{4.10}) -- (\ref{4.12})]
\begin{eqnarray}
 \biggr\{-[(\omega-(m+1)\Omega]^2+\kappa^2-4\Omega^2+&&k^2c_{\rm s}^2\biggr\}
          \breve{\xi}^{\rm int}_{r,+,\tilde{n}}
      -i2\Omega[\omega-(m+1)\Omega]\breve{\xi}^{\rm int}_{\varphi,+,\tilde{n}} \nonumber \\
      &&+i(kH)\Omega_\bot^2\breve{\xi}^{\rm int}_{z,+,\tilde{n}}  
      =\breve{A}_{r,+,\tilde{n}},
\label{5.8}
\end{eqnarray}
\begin{equation}
     -[\omega-(m+1)\Omega]^2\breve{\xi}^{\rm int}_{\varphi,+,\tilde{n}}
          +i2\Omega[\omega-(m+1)\Omega]\breve{\xi}^{\rm int}_{r,+,\tilde{n}}
       =\breve{A}_{\varphi,+,\tilde{n}},
\label{5.9}
\end{equation}
\begin{equation}
       \biggr\{-[(\omega-(m+1)\Omega]^2+\tilde{n}\Omega_\bot^2\biggr\}
                \breve{\xi}^{\rm int}_{z,+,\tilde{n}}
          -i\tilde{n}(kH)\Omega_\bot^2\breve{\xi}^{\rm int}_{r,+,\tilde{n}}
        =\breve{A}_{z,+,\tilde{n}},
\label{5.10}
\end{equation}
where $\tilde{n}=n+1$ or $n-1$ in case (i), and $\tilde{n}=n$ in case (ii).

In the case of intermediate oscillations characterized by 
$\hat{\mbox{\boldmath{$\xi$}}}_-^{\rm int}$, we can derive similar equations
as the above.
That is, we denote the right-hand side of equation (\ref{2.8}) by 
$\breve{\mbox{\boldmath{$A$}}}_-$ with similar expressions as equations 
(\ref{5.7}) -- (\ref{5.7''}). 
Then, as equations describing $\breve{\xi}_{r,-,\tilde{n}}$,
$\breve{\xi}_{\varphi,-,\tilde{n}}$,and $\breve{\xi}_{z,-,\tilde{n}}$,
we obtain the same equations as (\ref{5.8}) -- (\ref{5.10}), respectively,
except that $m+1$ is changed to $m-1$ and the subscript + in 
all variables are now changed to the subscript $-$.
                                 
The equations describing the intermediate oscillations, i.e., equations
(\ref{5.8}) -- (\ref{5.10}), and the corresponding equations for 
$\breve{\mbox{\boldmath{$\xi$}}}^{\rm int}_{-,\tilde{n}}$ can easily 
be solved with respect to $\breve{\xi}_{r,\pm,\tilde{n}}$,
$\breve{\xi}_{\varphi,\pm,\tilde{n}}$, and $\breve{\xi}_{r,\pm,\tilde{n}}$.
The results show that they are expressed in a form of $N_\pm/D_\pm$,
where $N_+$, for example,  is some combination of 
$\breve{\mbox{\boldmath $A$}}_{+, \tilde{n}}$'s on the right-hand side 
of equations (\ref{5.8}) -- (\ref{5.10}) and $D_\pm$ is
\begin{eqnarray}
   D_\pm(\omega,m\pm 1,\tilde{n})\equiv
      \biggr\{&&-[\omega-(m\pm 1)\Omega]^2+\kappa^2\biggr\}
      \biggr\{-[\omega-(m\pm 1)\Omega]^2+\tilde{n}\Omega_\bot^2\biggr\} \nonumber \\
       &&-[\omega-(m\pm 1)\Omega]^2k^2c_{\rm s}^2.
\label{5.11}
\end{eqnarray}

\subsection{Resonance}

The appearance of $D_\pm$ in the denominator comes from the fact that
the homogeneous wave equation for the intermediate oscillations gives the 
dispersion relation as the solvability condition 
and this is  $D_\pm(\omega,m\pm 1,\tilde{n})=0$.
In other word, at the radius where $D_\pm(\omega,m\pm 1,\tilde{n})=0$ is
satisfied, the intermediate oscillations resonantly response to the
external forces given by the right-hand side  of the wave equations.
In the case of intermediate oscillations of 
$\breve{\mbox{\boldmath $\xi$}}^{\rm int}_{+,\tilde{n}}$,
the external forcing terms are the right-hand sides of equations
(\ref{5.8}) -- (\ref{5.10}).

Since we are considering the cases where $k^2c_{\rm s}^2$ is much smaller
than $\Omega^2$ (i.e., geometrically thin disks),
the last term of the expression for $D_\pm$ can be approximately neglected
(particle approximation).
This means that the resonance occurs at two radii.
One occurs at a radius close to the radius where 
\begin{equation}
        [\omega-(m\pm 1)\Omega]^2-\kappa^2=0 \qquad(\rm {horizontal \ \ resonance})
\label{hresonance}
\end{equation}
holds, and the other does at a radius where 
\begin{equation}
   [\omega-(m\pm 1)\Omega]^2=\tilde{n}\Omega_\bot^2.
              \qquad (\rm {vertical \ \ resonance})
\label{vresonance}
\end{equation}
The former is called hereafter the horizontal resonance since the resonance occurs by 
horizontal motions,  and the latter the vertical resonance
since it occurs by vertical motions.
Our purpose here is to evaluate the right-hand side of equation (\ref{3.3})
and (\ref{3.4}) in these two cases.
In order to explicitly see the effects of resonance, we introduce variables
$\breve{\zeta}_{r,\pm, \tilde{n}}$,  $\breve{\zeta}_{\varphi,\pm, \tilde{n}}$,  
and $\breve{\zeta}_{z,\pm, \tilde{n}}$ defined by
\begin{equation}
      \breve{\xi}^{\rm int}_{r,\pm,\tilde{n}}=
          {\breve{\zeta}_{r,\pm,\tilde{n}}\over D_\pm},
\quad 
      \breve{\xi}^{\rm int}_{\varphi,\pm,\tilde{n}}=
          {\breve{\zeta}_{\varphi,\pm,\tilde{n}}\over D_\pm},
\quad
      \breve{\xi}^{\rm int}_{z,\pm,\tilde{n}}=
          {\breve{\zeta}_{z,\pm,\tilde{n}}\over D_\pm}.
\label{5.12}
\end{equation}
Then, considering that 
$({1/ 2})\rho_0[\mbox{\boldmath $C$}(\hat{\mbox{\boldmath $\xi$}}^*,
           \hat{\mbox{\boldmath $\xi$}}^{{\rm W}*})
         + \mbox{\boldmath $C$}(\hat{\mbox{\boldmath $\xi$}}^{{\rm W}*},
           \hat{\mbox{\boldmath $\xi$}}^{*})] $
is just the complex conjugate of the right-hand side of inhomogeneous
wave equation with respect to $\hat{\mbox{\boldmath $\xi$}}_+^{\rm int}$
[see equation (\ref{2.7})],
and that the inhomogeneous wave equations are explicitly written by
equations (\ref{5.8}) -- (\ref{5.10}), we have
\begin{eqnarray}
   {1\over  2}\rho_0\hat{\mbox{\boldmath $\xi$}}_+^{\rm int}\cdot
        && [\mbox{\boldmath $C$}(\hat{\mbox{\boldmath $\xi$}}^*,
           \hat{\mbox{\boldmath $\xi$}}^{{\rm W}*})
         + \mbox{\boldmath $C$}(\hat{\mbox{\boldmath $\xi$}}^{{\rm W}*},
           \hat{\mbox{\boldmath $\xi$}}^{*})] \nonumber \\ 
         =\rho_0{1\over D_+^2}&&
      \biggr[\biggr\{-[\omega-(m+1)\Omega]^2+\kappa^2-4\Omega^2+k^2c_{\rm s}^2\biggr\}
                 \breve{\zeta}_{r,+,\tilde{n}} \breve{\zeta}_{r,+,\tilde{n}}^* \nonumber \\
      && +i2\Omega[\omega-(m+1)\Omega]
                                 (\breve{\zeta}_{r,+,\tilde{n}} \breve{\zeta}_{\varphi,+,\tilde{n}}^*
                                  - \breve{\zeta}_{r,+,\tilde{n}}^* \breve{\zeta}_{\varphi,+,\tilde{n}})
                                               \nonumber \\
       && + i\tilde{n}k^2c_{\rm s}^2(-\breve{\zeta}_{r,+,\tilde{n}} \breve{\zeta}_{z,+,\tilde{n}}^*
                                                  + \breve{\zeta}_{r,+,\tilde{n}}^* \breve{\zeta}_{z,+,\tilde{n}})
                                                    \nonumber  \\
       && -[\omega-(m+1)\Omega]^2 \breve{\zeta}_{\varphi,+,\tilde{n}}
                                             \breve{\zeta}_{\varphi,+,\tilde{n}}^*
                       -\biggr\{[\omega-(m+1)\Omega]^2-\tilde{n}\Omega_\bot^2\biggr\} 
                                     \breve{\zeta}_{z,+,\tilde{n}} \breve{\zeta}_{z,+,\tilde{n}}^*\biggr].
\label{5.13}
\end{eqnarray}

Our purpose here is to evaluate the imaginary part of the volume integration of the
right-hand side of equation (\ref{5.13}) to have an expression for $W_+$.
As we see later, $D_\pm$ is proportional to $r-r_{\rm c}$ near the resonant radius
$r_{\rm c}$ of $D_\pm=0$.
Here, $r_{\rm c}$ is a function of $\omega$, $m$, $\tilde{n}$.
The terms in the brackets of equation (\ref{5.13}) are all real at a glance.
Hence, the radial integration of the right-hand side of equation (\ref{5.13}) has no imaginary 
part, unless the terms in the brackets have terms proportional to $D_+$.
If they have terms proportional to $D_+$, however, the right-hand side of equation (\ref{5.13})
has terms proportional to $1/D_+$ with real coefficients.
Since, $1/D_+$ is a pole in the complex $r$-plane,
the radial integration of $1/D_+$ brings 
about an imaginary part, as carefully discussed later.

In the case of the horizontal resonance, we have
\begin{equation}
     -[\omega-(m+1)\Omega]^2+\kappa^2+k^2c_{\rm s}^2
               \sim {D_+\over -\kappa^2+\tilde{n}\Omega_\bot^2},
\label{5.14}
\end{equation}
while in the case of the vertical resonance, we obtain
\begin{equation}
      -[\omega-(m+1)\Omega]^2+\tilde{n}\Omega_\bot^2
               \sim{D_+\over \kappa^2-\tilde{n}\Omega_\bot^2}.
\label{5.15}
\end{equation}
Under these considerations, we pick up only, from the 
right-hand side of equation (\ref{5.13}),
the terms which contribute to the imaginary 
part when they are integrated in the radial direction.
Then, we have  
\begin{eqnarray}  
       W_{{\rm H},+}&&={\omega\over 2}\Im \int{1\over 2}\rho_0\hat{\mbox{\boldmath $\xi$}}_+^{\rm int}
         [\mbox{\boldmath $C$}(\hat{\mbox{\boldmath $\xi$}}^*,
                  \hat{\mbox{\boldmath $\xi$}}^{{\rm W}*})
        +\mbox{\boldmath $C$}(\hat{\mbox{\boldmath $\xi$}}^{{\rm W}*},
                  \hat{\mbox{\boldmath $\xi$}}^*)]dV      
                          \nonumber    \\
           &&={\omega\over 2}\Im \int \rho_0{\cal H}_{\tilde{n}}^2\biggr({z\over H}\biggr)
                         {1\over D_{{\rm H},+}}\vert\breve{\zeta}_{r,+,\tilde{n}}\vert^2
                         {1\over -\kappa^2+\tilde{n}\Omega_\bot^2}dV
                           \nonumber \\
           &&={\omega\over 2}(2\pi)^{3/2}\tilde{n}!
                   \Im\int {1\over D_{{\rm H},+}}\vert\breve{\zeta}_{r,+,\tilde{n}}\vert^2
                         {r\rho_{00}(r)H\over -\kappa^2+\tilde{n}\Omega_\bot^2}dr
\label{5.16}
\end{eqnarray}
for the horizontal resonance.
The final equality is obtained by performing the vertical integration by using
the nature of the Hermite polynomials.
Here, the subscript H has been attached to $W_+$ and $D_+$ in order 
to emphasize that the horizontal resonance is now considered.
In the case of the vertical resonance, we have
\begin{eqnarray}  
       W_{{\rm V},+}&&={\omega\over 2}\Im \int \rho_0\hat{\mbox{\boldmath $\xi$}}_+^{\rm int}
         [\mbox{\boldmath $C$}(\hat{\mbox{\boldmath $\xi$}}^*,
                  \hat{\mbox{\boldmath $\xi$}}^{{\rm W}})
        +\mbox{\boldmath $C$}(\hat{\mbox{\boldmath $\xi$}}^{{\rm W}},
                  \hat{\mbox{\boldmath $\xi$}}^*)]dV      
                          \nonumber    \\
           &&={\omega\over 2}\Im \int \rho_0 {\cal H}_{\tilde{n}-1}^2\biggr({z\over H}\biggr)
                         {1\over D_{{\rm V},+}}\vert{\breve{\zeta}}_{z, +,\tilde{n}}\vert^2
                         {r\rho_{00}(r)H \over \kappa^2-\tilde{n}\Omega_\bot^2}dr
                             \nonumber \\
           &&={\omega\over 2}(2\pi)^{3/2}(\tilde{n}-1)!
                   \Im\int {1\over D_{{\rm V},+}}\vert\breve{\zeta}_{z,+,\tilde{n}}\vert^2
                         {r\rho_{00}(r)H\over \kappa^2-\tilde{n}\Omega_\bot^2}dr,
\label{5.17}
\end{eqnarray}
where the subscript V has been attached to $W_+$ and $D_+$ in order to
clearly show that the vertical resonance is considered.
It is noted that in the case of the vertical resonance, $W$, is proportional
to $({\tilde n}-1)!$, not ${\tilde n}!$.

In the case of coupling through $\hat{\mbox{\boldmath $\xi$}}_-^{\rm int}$,
we have similar expressions as the above.
They are
\begin{equation}  
       W_{{\rm H},-}={\omega\over 2}(2\pi)^{3/2}\tilde{n}!\Im \int 
                         {1\over D_{{\rm H},-}}\vert\breve{\zeta}_{r,-,\tilde{n}}\vert^2
                         {r\rho_{00}H\over -\kappa^2+\tilde{n}\Omega^2}dr
\label{5.18}
\end{equation} 
\begin{equation}  
       W_{{\rm V},-}={\omega\over 2}(2\pi)^{3/2}(\tilde{n}-1)!\Im \int 
                         {1\over D_{{\rm V},-}}\vert{\breve\zeta}_{z,-,\tilde{n}}\vert^2
                         {r\rho_{00}H\over \kappa^2-\tilde{n}\Omega^2}dr.
\label{5.19}
\end{equation}
It is noted that $\tilde{n}=n+1$ or $n-1$ in the case where the disk deformation is 
a warp [case (i)], while $\tilde{n}=n$ in the spiral deformation of disks which 
are symmetric with respect to the equatorial plane [case (ii)]. 

\subsection{Integrations of $1/D$ around the Resonant Radius}
     
The imaginary part of equations (\ref{5.16}) -- (\ref{5.19})
comes from the integration around the resonant radius, since the radius is a pole
of the integrant  in the complex $r$-plane.
To perform the integration around a pole, we must be careful.
If the frequency of oscillations, $\omega$, is taken to be real, the pole is just on
the real axis of $r$, and it is a problem what path we should take around the pole,
when we make integration along the real $r$-axis.
It is known in plasma physics, however, that the following procedure can satisfy the 
requirement of causality.
That is, we perform the integration along the real $r$-axis, tentatively assuming that the 
oscillations in consideration are growing 
(i.e., $\omega_{\rm i}<0$ in the present problem), and then the results 
obtained in this way are analytically extended to the case of $\omega_{\rm i}>0$.  
Hereafter, we examine the horizontal and vertical resonance 
separately with this standard procedure.

\noindent
(a) {\it Horizontal resonance}

For the horizontal resonance, we have $[\omega-(m\pm 1)\Omega]^2-\kappa^2
\sim 0$ and around the resonant radius, $r_{\rm c}$, the first term of the
Tayor expansion of $D_{{\rm H},\pm}$ gives
\begin{equation}
       D_{{\rm H},\pm}=2(\tilde{n}\Omega_\bot^2-\kappa^2)G_{{\rm H},\pm}
             \biggr[{r-r_{\rm c}\over r_{\rm c}}-i{\omega-(m\pm 1)\Omega_{\rm c}
                    \over G_{{\rm H},\pm}}\omega_{\rm i}\biggr],
\label{5.20}
\end{equation}
where 
\begin{equation}
      G_{{\rm H},\pm}=\biggr\{(m\pm 1)\Omega[\omega-(m\pm 1)\Omega]
            {d{\rm ln}\Omega\over d{\rm ln}r}
           +\kappa^2{d{\rm ln}\kappa\over d{\rm ln}r}\biggr\}_{\rm c},
\label{5.21}
\end{equation}
and the subscript c denotes the values at the resonant radius.
Considering these expressions and assuming $\omega_{\rm i}<0$ 
(the requirement of causality), we have (Kato 2004)
\begin{equation}
       \int {dr\over D_{{\rm H},\pm}}=
        -i{\pi r_{\rm c}\over 2(\tilde{n}\Omega_\bot^2-\kappa^2)_{\rm c}
                       \vert G_{{\rm H},\pm}\vert}
          {\rm sign}[\omega-(m\pm 1)\Omega]_{\rm c}.
\label{5.22}
\end{equation}
It is noted that this result is valid even when $\omega_{\rm i}>0$ as mentioned before,
although it is obtained under the assumption of $\omega_{\rm i}<0$.

Combining equations (\ref{5.16}) [or  (\ref{5.18})]  and (\ref{5.22}),
we have, for horizontal resonance,
\begin{equation}
    W_{{\rm H}, \pm}=-{\omega\over 4} (2\pi)^{3/2}\pi\tilde {n}!
         \biggr[{r^2\rho_{00}H\over (\tilde{n}\Omega_\bot^2-\kappa^2)^2}\biggr]_{\rm c}
        {\vert\breve{\zeta}_{r,\pm,\tilde{n}}\vert_{\rm c}^2 \over \vert G_{{\rm H},\pm}\vert}
            {\rm sign}[\omega-(m\pm 1)\Omega]_{\rm c}.
\label{5.23}
\end{equation}
This result shows that the sign of the work done on the oscillations by the horizontal 
resonance is determined by the sign of $\omega-(m\pm 1)\Omega$ at the
resonant radius.
The work is positive if $\omega-(m\pm 1)\Omega$ at the resonant radius is negative.

\noindent
(b) {\it Vertical resonance}

The work done on the oscillations by the vertical resonance is obtained by a similar
way discussed above.
That is, in the case of vertical resonance, the resonance occurs at a radius close to
$[\omega-(m\pm 1)\Omega]^2-\tilde{n}\Omega_\bot^2=0$.
Around this resonant radius, $r_{\rm c}$, we have
\begin{equation}
       D_{{\rm V},\pm}=-2(\tilde{n}\Omega_\bot^2-\kappa^2)G_{{\rm V},\pm}
             \biggr[{r-r_{\rm c}\over r_{\rm c}}-i{\omega-(m\pm 1)\Omega_{\rm c}
                    \over G_{{\rm V},\pm}}\omega_{\rm i}\biggr],
\label{5.24}
\end{equation}
where
\begin{equation}
      G_{{\rm V},\pm}=\biggr\{(m\pm 1)\Omega[\omega-(m\pm 1)\Omega]
            {d{\rm ln}\Omega\over d{\rm ln}r}
           +2\tilde{n}\biggr(\Omega_\bot{d\Omega_\bot\over d{\rm ln}r}\biggr)
                 \biggr\}_{\rm c}.
\label{5.25}
\end{equation}
The radial integration of $1/D_{\rm V,\pm}$ along the real $r$-axis then gives (Kato 2004)
\begin{equation}
       \int {dr\over D_{{\rm V},\pm}}=
          i{\pi r_{\rm c}\over 2(\tilde{n}\Omega_\bot^2-\kappa^2)_{\rm c}
                       \vert G_{{\rm V},\pm}\vert}
          {\rm sign}[\omega-(m\pm 1)\Omega]_{\rm c}.
\label{5.26}
\end{equation}
Then, from equations (\ref{5.17}) [or (\ref{5.19})] and (\ref{5.26}), we have, 
for the vertical resonance,
\begin{equation}
    W_{{\rm V},\pm}=-{\omega\over 4} (2\pi)^{3/2}\pi(\tilde {n}-1)!
         \biggr[{r^2\rho_{00}H\over (\tilde{n}\Omega_\bot^2-\kappa^2)^2}\biggr]_{\rm c}
        {\vert\breve{\zeta}_{z,\pm,\tilde{n}}\vert_{\rm c}^2 \over \vert G_{{\rm V},\pm}\vert}
            {\rm sign}[\omega-(m\pm 1)\Omega]_{\rm c}.
\label{5.27}
\end{equation}
It is noted that in the vertical resonance the sign of work done on oscillations 
is also determined by the sigh of $\omega-(m\pm 1)\Omega$ 
at the resonant radius.
The radius of resonance is, however, different from that in the 
case of the horizontal resonance, although both radii of the horizontal and
vertical resonances are simply denoted by the same symbol $r_{\rm c}$, without 
distinguishing them. 

\subsection {Stability Condition}

The wave energy given by equation (\ref{2.13}) can be expressed as
\begin{equation}
    E={(2\pi)^{3/2}\over 2}\omega(\omega-m\Omega)_{\rm c}(r^4H\rho_{00})_{\rm c}E_n,
\label{5.28}
\end{equation}
where $E_n$ is a dimensionless quantity given by
\begin{equation}
    E_n=\int {rH\rho_{00}\over (rH\rho_{00})_{\rm c}}{\omega-m\Omega 
                                           \over(\omega-m\Omega)_{\rm c}}
        \biggr(n!{\vert \breve{\xi}_r\vert ^2\over r_{\rm c}^2}
        +(n-1)!{\vert \breve{\xi}_z\vert ^2\over r_{\rm c}^2}\biggr){dr\over r_{\rm c}}.
\label{5.29}
\end{equation}
Hence, the growth rate, $-\omega_{\rm i}$ [see equation (\ref{2.17}) or (\ref{2.19})], can be expressed
in the case of the horizontal resonance as
\begin{equation}
   -\omega_{{\rm i},{\rm H}, \pm}=-{\pi\over 4}{\tilde n}!\biggr[{(\breve{\zeta}_{r,\pm,\tilde{n}}/r)^2\over
                         (\omega-m\Omega)(\tilde{n}\Omega_\bot^2-\kappa^2)^2\vert G_{{\rm H},\pm}\vert}
                                 \biggr]_{\rm c}
          {{\rm sign}[\omega-(m\pm 1)\Omega]_{\rm c}\over E_n}.
\label{5.30}
\end{equation}
Similarly, in the case of vertical resonance, we have from equation (\ref{5.27})
\begin{equation}
   -\omega_{{\rm i}, {\rm V},\pm}=-{\pi\over 4}({\tilde n}-1)!\biggr[{(\breve{\zeta}_{z,\pm,\tilde{n}}/r)^2\over
                         (\omega-m\Omega)(\tilde{n}\Omega_\bot^2-\kappa^2)^2\vert G_{{\rm H},\pm}\vert}
                                 \biggr]_{\rm c}
          {{\rm sign}[\omega-(m\pm 1)\Omega]_{\rm c}\over E_n}.
\label{5.31}
\end{equation}
The growth rates given by equations (\ref{5.30}) and (\ref{5.31}) show that 
the resonance leads to
growth of oscillations when the sign of $[\omega-(m\pm 1)\Omega]/(\omega-m\Omega)$
at the resonant radius is negative.
In the followings, we examine in what types of resonance and in what wave modes the sign becomes negative.

\section{Resonant Radii and Resonant Frequencies of Locally Unstable Oscillations}

This examination is separately made below for various combinations of
types of resonance and of wave mode.

\subsection{Horizontal Resonance of Inertial-Acoustic or G-Mode Oscillations}

In the horizontal resonance, the resonance is realized at the radius where 
$[\omega-(m\pm 1)\Omega]^2-\kappa^2\sim 0$ [see equation (\ref{hresonance})].
The inertial-acoustic or g-mode oscillations will be predominantly exist at the 
radius where $(\omega-m\Omega)^2-\kappa^2\sim 0$ is satisfied (see e.g., Kato and
Fukue 2006).
For resonance to occur efficiently, both radii of $[\omega-(m\pm 1)\Omega]^2=\kappa^2$
and $(\omega-m\Omega)^2=\kappa^2$ must be the same.
This condition is satisfied at the radius of
\begin{equation}
      \kappa=\frac{\Omega}{2}
\label{5.32}
\end{equation}
in two cases where 
\begin{eqnarray}
    &&{\rm case\ \  (a)}:    
             \omega-(m+1)\Omega=-\kappa \quad {\rm  and}\quad \omega-m\Omega=\kappa
                                                       \nonumber \\
    &&{\rm case\ \ (b)}:  \omega-(m-1)\Omega=\kappa \ \ \quad {\rm and}\quad  
                                                                   \omega-m\Omega=-\kappa.
\label{5.32'}
\end{eqnarray}
We see that in both cases the sign of $[\omega-(m+1)\Omega]/(\omega-m\Omega)$
or $[\omega-(m-1)\Omega]/(\omega-m\Omega)$ is negative.
This implies that $-\omega_{\rm i}>0$ and the resonance amplifies the oscillations.

Let us see more in detail.
In the former case of (a), $[\omega-(m+1)\Omega]_{\rm c}<0$ and
the work done on the oscillations is positive [see equation (\ref{5.23})], and
the wave energy is positive [see equation (\ref{5.28})].
Hence, the oscillations grow.
In the latter case of (b), $[\omega-(m-1)\Omega]_{\rm c}>0$
and a negative work is done on the oscillations.
The oscillations, however, grow since the wave energy is negative.
Various quantities related to resonance and amplification are summarized in
tables 1 and 2.
 
The above argument can be applied not only to the case where the disk is warped
(i.e., $m^{\rm W}=1$ and $n^{\rm W}=1$), but also 
to the case where the disk is deformed by a one-armed pattern which is symmetric 
with respect to the equatorial plane (i.e., $m^{\rm W}=1$ and $n^{\rm W}=0$).

\subsection{Horizontal Resonance of C-Mode or Vertical P-Mode Oscillations}

These oscillation modes exist predominantly at the radius where
$(\omega-m\Omega)^2-n\Omega_\bot^2\sim 0$.
Both conditions of  $[\omega-(m\pm 1)\Omega]^2=\kappa^2$ and
$(\omega-m\Omega)^2=n\Omega_\bot^2$ can be simultaneously satisfied 
at the radius of 
\begin{equation}
    \kappa=n^{1/2}\Omega_\bot -\Omega.
\label{5.33}
\end{equation}
when
\begin{eqnarray}
   &&{\rm case\ \ (a)}: \omega-(m+1)\Omega=\kappa\quad\ \  {\rm  and}
                                \quad \omega-m\Omega=n^{1/2}\Omega_\bot   \nonumber \\
   && {\rm case\ \ (b)}: \omega-(m-1)\Omega=-\kappa \quad {\rm and}
                                \quad \omega-m\Omega=-n^{1/2}\Omega_\bot.
\label{5.33'}
\end{eqnarray}
This consideration shows that at the resonant point the sign of
$[\omega-(m\pm 1)\Omega]/(\omega-m\Omega)$ is positive, showing
that $-\omega_{{\rm i}, {\rm H},\pm}$ given by equation (\ref{5.30}) is negative.
That is, the resonance dampens the oscillations (see tables 1 and 2).

\subsection{Vertical Resonance of Inertial-Acoustic or G-Mode Oscillations}

The condition of the vertical resonance is $[\omega-(m\pm 1)\Omega]^2
\sim \tilde {n}\Omega_\bot^2$.
The radius where the inertial-acoustic or g-mode oscillations 
predominantly exists is $(\omega-m\Omega)^2=\kappa^2$.
Combination of these two relations shows that the resonance occurs at
\begin{equation}
    \kappa=\tilde{n}^{1/2}\Omega_\bot-\Omega,
\label{5.34}
\end{equation}
when 
\begin{eqnarray}
     && {\rm case \ (a)}:  \omega-(m-1)\Omega={\tilde{n}}^{1/2}\Omega_\bot 
                         \quad\quad {\rm and} \quad
                          \omega-m\Omega=\kappa   \nonumber \\
      && {\rm case \ (b)}:
                   \omega-(m+1)\Omega=-{\tilde{n}}^{1/2}\Omega_\bot \quad {\rm and}
                           \quad \omega-m\Omega=-\kappa.
\label{5.34'}
\end{eqnarray}
In these two cases the sigh of $[\omega-(m\pm 1)\Omega]/(\omega-m\Omega)$ is
positive.
Hence, equations (\ref{5.31}) shows that the resonance dampens the oscillations.

The vertical resonance of c- and vertical p-mode oscillations is absent, since
the conditions of $(\omega-m\Omega)^2-{\tilde{n}}^{1/2}\Omega_\bot=0$
and $(\omega-m\Omega)^2-n^{1/2}\Omega_\bot=0$ can not be simultaneously
satisfied at any radius.

\subsection{Rough Estimate of Growth Rate}

Finally, the order of growth rate is briefly estimated in the case of horizontal
resonance of inertial-acoustic or g-mode oscillations in warped disks.
To do so, we must evaluate the order of $\breve{\xi}_r^{\rm int}$.
The nonlinear coupling terms between $\mbox{\boldmath $\xi$}$ and
$\mbox{\boldmath $\xi$}^{\rm W}$ are given by equations (\ref{2.5}) and 
(\ref{2.6}).
Some detailed calculations show that the radial component of 
$\rho_0\mbox{\boldmath $C$}_{\rm p}$, for example, is on the order of 
$\rho_0\xi_r\eta\Omega^2(kH)/\lambda^{\rm W}$ (Kato 2004), where
$\xi_r$ and $\eta$ are the orders of the radial displacement associated with
the oscillations and of the vertical displacement associated with the warp, respectively,
and $1/k$ and $\lambda^{\rm W}$ are the radial wavelengths of the disk oscillation
and the warp, respectively.
Then, the wave equations describing the intermediate oscillations give the order of
$\breve{\xi}_r^{\rm int}$ as 
\begin{equation}
   {\cal O}(\breve{\xi}_r^{\rm int})={\xi_r\eta (kH) \over \lambda^{\rm W}}.
\label{5.35}
\end{equation}
Then, we have 
\begin{equation}
             {\cal O}(\breve{\zeta}_r^{\rm int}) 
                      =\xi_r\eta\Omega^4{kH\over \lambda^{\rm W}}.
\label{5.36}
\end{equation}
The dimensionless wave energy contained in the disk, $E_n$, is found to be [see equation
(\ref{5.29})]
\begin{equation}
              {\cal  O}(E_n)=\biggr(\frac{\xi_r}{r_{\rm c}}\biggr)^2{L\over r_{\rm c}},
\label{5.37}
\end{equation}
where $L$ is the radial extend where the oscillations exist.
Using these relations, we obtain from equation (\ref{5.30}) 
that the growth rate is on the order of 
\begin{equation}
         {\cal O}(-\omega_{\rm i})=\alpha(kH)^2\Omega,
\label{5.38}
\end{equation}
where $\alpha=(\eta/\lambda^{\rm W})^2(r_{\rm c}/L)$,
confirming the results obtained by Kato (2004).\footnote{
In Kato (2004), $\lambda^{\rm W}$ is taken to be $r_{\rm c}$.
}

\begin{longtable}{cccccc}
\caption{Resonant oscillations in warped disks ($m^{\rm W}=1$, $n^{\rm W}=1$).}
\label{tab:1}
\endfirsthead
\hline\hline
  \multicolumn{6}{c}
       {Inertial-acoustic mode and g-mode oscillations} \\
  \hline
    type of resonance & type of coupling & resonant radius & work 
                                & wave energy & stability \\
\hline
    horizontal  & $m$ $\rightarrow$ $m+1$ $\rightarrow$ $m$  & $r/r_{\rm g}=4.0$ 
                     & positive & positive & {\bf growth} \\
                     & $m$ $\rightarrow$ $m-1$ $\rightarrow$ $m$   & $r/r_{\rm g}=4.0$
                     & negative & negative & {\bf growth} \\
\cline{2-6}
   vertical       &  $m$ $\rightarrow$ $m+1$ $\rightarrow$ $m$  & $r/r_{\rm g}=3.62,\  6.46$ 
                     & negative & positive & damping \\
                      & $m$ $\rightarrow$ $m-1$ $\rightarrow$ $m$   & $r/r_{\rm g}=3.62,\  6.46$
                     & positive & negative & damping \\
\hline\hline
\multicolumn{6}{c}
     {c-mode and vertical p-mode oscillations} \\ 
  \hline
    type of resonance & type of coupling & resonant radius & work 
                                & wave energy & stability \\
\hline
   horizontal  & $m$ $\rightarrow$ $m+1$ $\rightarrow$ $m$  & $r/r_{\rm g}=3.62,\ 6.46$ 
                     & negative & positive & damping \\
                     & $m$ $\rightarrow$ $m-1$ $\rightarrow$ $m$   & $r/r_{\rm g}=3.62,\ 6.46$
                     & positive & negative & damping \\
\cline{2-6}
     vertical    &   \multicolumn{5}{c} {absence} \\
\hline
\end{longtable}

\begin{longtable}{cccccc}
\caption{Resonant oscillations deformed by one-armed perturbations 
($m^{\rm W}=1$, $n^{\rm W}=0$).}
\label{tab:2}
\endfirsthead
\hline\hline
  \multicolumn{6}{c}
       {Inertial-acoustic mode and g-mode oscillations}\\
\hline
    type of resonance & type of coupling & resonant radius & work 
                                & wave energy & stability \\
\hline
    horizontal  & $m$ $\rightarrow$ $m+1$ $\rightarrow$ $m$  & $r/r_{\rm g}=4.0$ 
                     & positive & positive & {\bf growth} \\
                     & $m$ $\rightarrow$ $m-1$ $\rightarrow$ $m$   & $r/r_{\rm g}=4.0$
                     & negative & negative & {\bf growth} \\
\cline{2-6}
     vertical       &  \multicolumn{5}{c} {absence}  \\
\hline\hline
\multicolumn{6}{c}
     {c-mode and vertical p-mode oscillations} \\
\hline
    type of resonance & type of coupling & resonant radius & work 
                                & wave energy & stability \\
\hline
   horizontal  & \multicolumn{5}{c} {absence}  \\
\cline{2-6}
     vertical    &   \multicolumn{5}{c} {absence} \\
\hline
\end{longtable}

\section{Resonant Excitation of Global Oscillations}

So far, we have considered resonant excitation of local oscillations, assuming 
that the oscillations are local in the radial direction in the sense that their 
radial wavelength is shorter than the characteristic radial 
scale-length of the disk (i.e., $kr>1$).
Certainly, we have extensively used the approximation in section 5 
when the work done on oscillations is calculated.
The processes of the analyses, however, suggest that the final result
(i.e., the sign of the work done on oscillations 
by resonance is governed by the sign of $\omega-(m\pm 1)\Omega$ at 
the resonant radius [see equations (63) and (67)])
is free from the approximation,
although the magnitude of the work depend on the radial form of oscillations. 
This anticipation seems to be supported by considering the other limit of
the particle approximation ($kr>1$), although it is not proved.

Different from the above, the resonant radii and resonant frequencies 
estimated in section 6 come from the approximation that the oscillations 
are local.\footnote{
In section 6 we have assumed that the oscillations exist predominantly 
around the boundary of the propagation region, 
since the group velocity of the local oscillations vanishes there.
}
Thus, in this section we relax the approximation
and examine how the results 
in section 6 are modified in cases of global oscillations.

We restrict, however, our attention only to the case where oscillations are 
inertial-acoustic or g-mode oscillations and resonance is horizontal, since 
we can show that in other cases oscillations are damped or the resonance 
appears only in the evanescent region of the oscillations.

\subsection{Horizontal Resonance of Inertial-Acoustic Oscillations}

Let us consider inertial-acoustic oscillations of frequency $\omega$ with 
azimuthal wavenumber $m$.
The propagation region of such inertial-acoustic oscillations is specified by
$(\omega-m\Omega)^2>\kappa^2$.
In other words, the evanescent region of the oscillations is 
$(\omega-m\Omega)^2<\kappa^2$, i.e., $m\Omega-\kappa< \omega
<m\Omega+\kappa$.
The evanescent regions in the $\omega$--$r$ plane (i.e., the so-called 
propagation diagram) 
are shown by dark area in figures 1 and 2 for $m=1$ for $m=2$, respectively.
The region in the case of $m=0$ is supplementarily shown in figure 3.

Next, we consider where the horizontal resonance occurs on the 
$\omega$--$r$ plane.
In the coupling through $m$ $\rightarrow$ $m+1$ $\rightarrow$ $m$, the 
resonance occurs on the curve of 
$[\omega-(m+1)\Omega]^2=\kappa^2$, while it occurs on the 
curve of $[\omega-(m-1)\Omega]^2=\kappa^2$ in the coupling through 
$m$ $\rightarrow$ $m-1$ $\rightarrow$ $m$ [see subsection 5.2].
These curves are shown in figures 1 to 3 
by thick (solid or dashed) or less thick dashed curves (see below for 
distinction of these curves).

The first problem to be examined is in which part of the curve of 
$[\omega-(m\pm 1)\Omega]^2=\kappa^2$, resonance excites or 
dampens the oscillations.
This can be examined by studying the sign of  
$[\omega-(m\pm 1)\Omega]/(\omega-m\Omega)$ on the curve
[see that $W$'s are given by equations (63) and (67) and the wave energy is 
given by equation (13)].
Above the curve of $\omega=m\Omega$ on the 
$\omega$--$r$ plane, the wave energy, $E$, is positive, 
while it is negative below the curve.
Next, in the case of coupling through $m$ $\rightarrow$ $m+1$
$\rightarrow$ $m$, the sign of the work done on oscillations is 
governed by the sign of $\omega-(m+1)\Omega$.
In the case of coupling through $m$ $\rightarrow$ $m-1$ $\rightarrow$
$m$, on the other hand, the sign of the work is governed by the sigh of 
$\omega-(m-1)\Omega$.
Considering these situations, we find that resonance excites (amplifies) 
oscillations in the case where it occurs on the thick parts (including thick dashed parts) of the curves of $[\omega-(m\pm 1)\Omega]^2=\kappa^2$.
However, resonance leads to damping of oscillations in the case where the 
resonant point is on less thick dashed parts of the curves.
These results are shown in figures 1 and 2 for $m=1$ and $m=2$, respectively,
and supplementarily in figure 3 for $m=0$.

Even when resonance excites oscillations, a strong excitation of the oscillations
will not be expected in the case where the resonant point occurs in the 
evanescent region of the oscillations.
This is because the coupling between
the oscillations and the disk deformation is weak due to small amplitude of
oscillations in the evanescent region.
Hence, the parts of the curves of $[\omega-(m\pm 1)\Omega]^2=\kappa^2$ 
where resonance excites oscillations but the resonance occurs in 
the evanescent region of oscillations are shown by thick and dashed curves.

As shown in figures 1 to 3, the thick solid parts of the resonant curves are
only in a finite range of frequency.
Propagations of some typical growing oscillations are shown in figures by horizontal 
dashed lines.
The resonant radius of these oscillations is shown by a large open circle on the lines.
It is noted that the cases considered in section 6 under the local approximations
are those where resonance is assumed to occur 
at the crossing point of the resonant curve and the boundary curve of the 
propagation region, i.e., at $4r_{\rm g}$.

The oscillations labeled by $\omega_{\rm LL}$ in figure 1 and by
$\omega_{\rm L}$ in figure 2 are trapped in a finite range in radius,
bounded between the inner edge of the disk ($\sim 3r_{\rm g}$) and
the outer barrier of $\omega=\Omega-\kappa$ (for $\omega_{\rm LL}$) or
of $\omega=2\Omega-\kappa$ (for $\omega_{\rm L}$).
Compared with these oscillations, the oscillations labeled by $\omega_{\rm H}$
(figure 1) and $\omega_{\rm HH}$ (figure 2) are not bounded in their outer 
boundary.
In this sense, the latter two oscillations will be less observable with definite
frequencies, compared with the former two ones.
This consideration suggests that the pair QPOs observed in some 
X-ray binaries are the oscillations labeled by $\omega_{\rm LL}$ and
$\omega_{\rm L}$ (see, e.g., Kato and Fukue 2006).  

It is noted that in the limit of local perturbations, the oscillations labeled by $\omega_{\rm LL}$ (figure 1) and
$\omega_{\rm L}$ (figure 2) tend to the oscillations of case (b) of
equation (73). 
Similarly, the oscillations labeled by $\omega_{\rm H}$ (figure 1) and $\omega_{\rm HH}$ (figure 2) tend to the oscillations of case (a) of 
equation (73).

Finally, resonant excitation of axisymmmetric ($m=0$) inertial-acoustic
oscillations is briefly mentioned.\footnote{
So far, we less discussed the axisymmetric oscillations by assuming that
they will not produce much light variation.
}
The results are shown in figure 3.
In the case of $m=0$, the resonance curve on the $\omega$--$r$ plane is $\omega=\Omega\pm\kappa$,
and the evanescent region of the oscillations is the dark area below the curve
of $\kappa(r)$.
The same considerations as those in cases of $m=1$ and $m=2$ show that
axisymmetric inertial-acoustic oscillations are excited in the case where 
resonance occurs on the curve of $\omega=\Omega-\kappa$.
(On the thick dashed part of the curve, resonance occurs only in the
evanescent region of the oscillations).
The resonance on the curve of $\omega=\Omega+\kappa$
dampens oscillations (less thick dashed curve).
The oscillations that can be excited by resonance are not trapped in a finite
region.
On the other hand, 
the oscillations whose frequencies are less than $\kappa_{\rm max}$
can be trapped, bounded inside by the inner edge of the disk and outside by the barrier of  $\omega=\kappa$. 
Such oscillations, however, do not have resonance in their propagation region.
  
 \begin{figure}
 \begin{center}
  \FigureFile(80mm,80mm){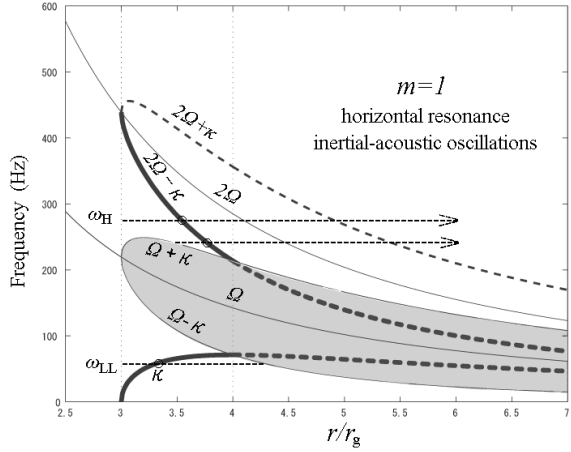}
 \end{center}
  \caption{
Frequency--radius relation, $\omega=(m\pm 1)\Omega\pm \kappa$
(four combinations of $\pm$ being possible), showing resonant radii
where inertial-acoustic oscillations with frequency $\omega$ and 
azimuthal wavenumber $m=1$ have resonant interaction with disk deformation
(thick solid, thick dashed, and less thick dashed curves).
Horizontal resonance of inertial-acoustic oscillations is considered.
In some part of the frequency-radius curve of $\omega=(m\pm 1)\pm\kappa$,
the resonance excites inertial-acoustic oscillations,
while in the other part the resonance dampens the oscillations.
The part of excitation is shown by thick solid (or thick dashed) curves, 
while the part of damping is shown by a less thick dashed curve.
Even in cases of excitation of oscillations, there are cases where the excitation 
occur in the evanescent region of the oscillations.
Such cases are shown by thick but dashed curves, since the
interaction between the oscillations and the deformation of disks is weak there.
The evanescent region of $m=1$ inertial-acoustic oscillations is shown as
shaded area.
For comparison, the curves of $\Omega(r)$ and $2\Omega(r)$
are also shown.
The horizontal dashed curves demonstrate the propagation 
of typical oscillations that are excited.
The open circles on the horizontal dashed curves show the radii where
resonance occurs.
The propagation of oscillations labeled by $\omega_{\rm LL}$
is bounded in a finite range: The inner boundary is the inner edge of
the disk and the outer boundary is the barrier of $\Omega-\kappa$.
In the case of oscillations labeled by $\omega_{\rm H}$, the 
propagation region is bounded only in the inner boundary.
In this sense, the latter oscillations labeled by $\omega_{\rm H}$
will be less observable with a definite frequency, compared with
the former oscillations of $\omega_{\rm LL}$.  
}
\end{figure}

\begin{figure}
  \begin{center}
 \FigureFile(80mm,80mm){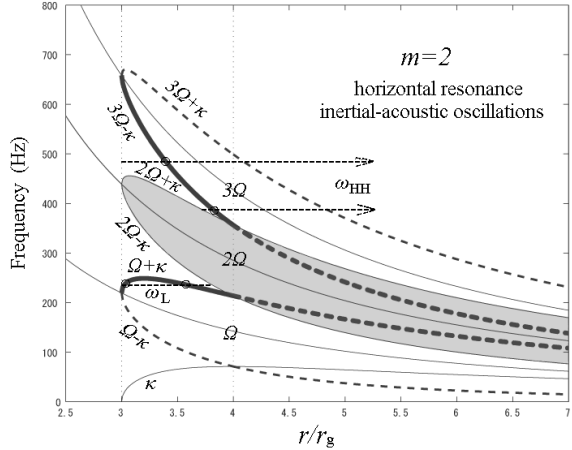}
  \end{center}
  \caption{
 The same as figure 1 except for $m=2$.
Because of the difference of $m$, the frequency--radius relation of
resonance is now $\omega=\Omega\pm\kappa$ and
$\omega=3\Omega\pm \kappa$ (both signs of $\pm$ being possible).
On the thick solid parts of the curves, resonance
excites oscillations in their propagation region, but on the thick dashed part  
resonance occurs in the evanescent region of oscillations.
The less thick dashed parts of the curves show the resonance that dampens 
inertial-acoustic oscillations. 
The evanescent region of the oscillations are now specified by
$2\Omega-\kappa<\omega<2\Omega+\kappa$ (dark area).   
For comparison, the curves of $\Omega$, $2\Omega$, $3\Omega$ are
also shown.
The propagation region of oscillations labeled by $\omega_{\rm L}$ is
bounded in a finite region.
However, the region of oscillations labeled by $\omega_{\rm HH}$ is not
bounded in the outer region.
In this sense, the latter oscillations will be less observable with a
definite frequency, compared with the former oscillations with $\omega_{\rm L}$.
}
\end{figure}

\begin{figure}
  \begin{center}
 \FigureFile(80mm,80mm){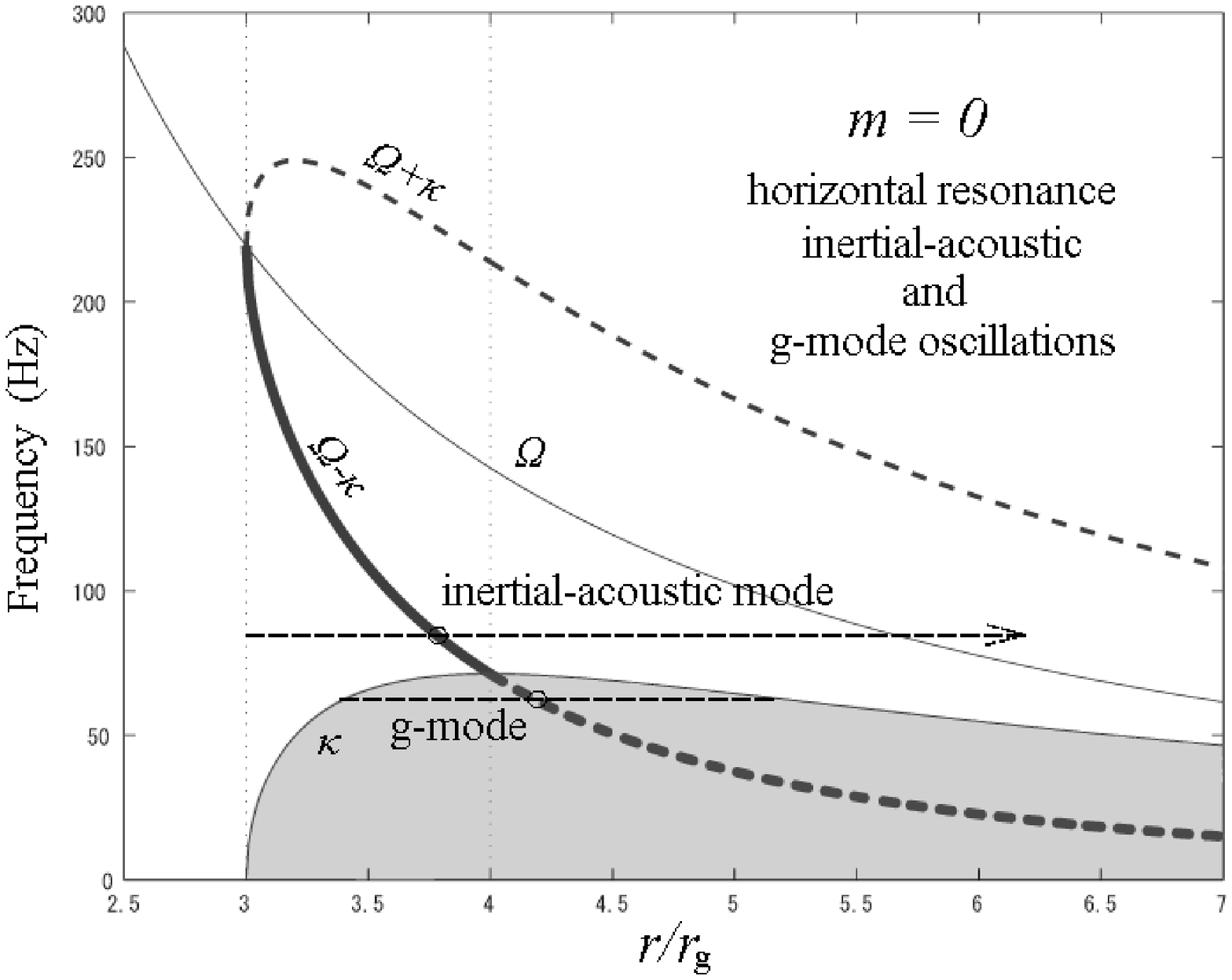}
  \end{center}
  \caption{
The same as figure 1 except for $m=0$.
The dark area represents the evanescent region of inertial-acoustic
oscillations (the propagation region of gravity oscillations). 
Oscillations with frequency of $\omega$ have resonance with disks
at the radius where $\omega=\Omega\pm \kappa$ is satisfied.
This frequency--radius relation is shown by separating into three parts;
thick solid, thick dahsed, and less thick dashed parts.
On the thick solid and thick dashed parts, resonance excites inertial-acoustic oscillations in their propagations region and in their evanescent region, 
respectively.
(In the case of g-mode oscillations, the meaning of the thick solid part and
the thick dashed one is exchanged.)
On the less thick dashed part, resonance dampens inertial-acoustic
oscillations.
It is noted that trapped g-mode oscillations whose frequencies are around 
$\kappa_{\rm max}$ can be excited by our present resonant model.
}
\end{figure}

\subsection{Horizontal Resonance of G-mode Oscillations}

The frequency--radius relation where resonance occurs for g-mode
oscillations with frequency $\omega$ and azimuthal wavenumber $m$ 
is the same as that for inertial-acoustic oscillations with 
the same $\omega$ and $m$.
On the curve of the frequency-radius relation, the part where resonance
dampens g-mode oscillations is also the same as that in the case of inertial-acoustic oscillations.

A difference between inertial-acoustic oscillations and g-mode ones
appears in their propagation regions. 
The propagation region of g-mode oscillations with frequency $\omega$
and azimuthal wavenumber $m$ 
is $(\omega-m\Omega)^2<\kappa^2$.
That is, the evanescent regions of the inertial-acoustic oscillations 
(the shaded areas in figures 1, 2 and 3) are now
the propagation regions of g-mode oscillations.
That is, the shaded region in the figure is now the propagation region, 
and thus the thick solid curve and the thick dashed one are interchanged.

This consideration shows that g-mode oscillations are trapped in a 
region with a finite radial extend, and in their propagation
region, a resonance occurs so as to amplify the oscillations 
(see figures 1, 2 and 3).
In the case of non-axisymmetric g-mode oscillations, however, 
we should remember the following situations.  
Except for the cases where the frequencies of g-mode oscillations are as high
as the upper value limited by the propagation region of the oscillations,
the corotation radius ($\omega=m\Omega$) appears in the propagation region
of the oscillations (see figures 1 and 2).
Different from the case of inertial-acoustic oscillations of $n=0$, 
the presence of corotation resonance in the propagation region strongly 
dampens the g-mode oscillations (Kato 2003; Li et al. 2003).
This strong damping suppresses the growth by the resonance by disk deformation 
considered here.
In a narrow frequency range where the frequencies are as high as the upper 
limit, the g-mode oscillations can be trapped in a radially narrow region 
without having corotation resonance in their propagation region
(cf. figure 1).
In this case, however, a resonance does not occur inside the propagation 
region (see figure 1).

To conclude, excitation (amplification) of global non-axisymmetric
g-mode oscillations is less promising compared with that of global 
non-axisymmetric inertial-acoustic oscillations.
Axisymmetric g-mode oscillations are, however, a good candidate
of one of observed quasi-periodic oscillations.
Axisymmetric g-mode oscillations are trapped around $\kappa_{\rm max}$ 
and excited,  since a resonance that can excite the oscillations appears in 
the propagation region of the oscillations (see figure 3).
This will be one of possible excitation mechanisms of the stable 67 Hz 
oscillation observed in GRS 1915+105.

\section{Discussion}

In this paper we have examined resonant excitation (amplification) of disk
oscillations in deformed disks.
Analyses are made in a general way as far as possible so that we can perspective
see the essence of the resonant amplification processes.

In a previous paper (Kato 2004) we have considered a warp as a possible disk 
deformation to excite oscillations.
In this paper, we show that, in addition to warps, a one-armed pattern symmetric 
with respect to the equatorial plane can also excite (amplify) disk oscillations.

As shown in the text, the horizontal resonance is of importance, since it can 
excite inertial-acoustic and g-mode oscillations.
The resonance occurs at the radii of the resonant condition,
$[\omega-(m\pm 1)\Omega]^2=\kappa^2$, being satisfied, where
$\omega$ and $m$ are frequency and azimuthal wavenumber of inertial-acoustic
(or g-mode) oscillations, respectively.

Important quantities determining the stability are,  (i) the sign of work done 
by deformed disks to disk oscillations through resonance at resonant radius, 
and (ii) the sign of wave energy.
As discussed in the text, the sign of the work done on the oscillations with
$\omega$ and $m$ is
determined by the sign of $\omega-(m+1)\Omega$ [or $\omega-(m-1)\Omega$] at
the resonant radius, depending that the path of coupling is $m$ $\rightarrow$ 
$m+1$ $\rightarrow$ $m$ or  $m$ $\rightarrow$ $m-1$ $\rightarrow$ $m$. 
The sign of wave energy is determined where the corotation radius exists.
If a wave is mainly inside the corotation radius of $\omega-m\Omega=0$,
the wave energy is negative, while it is positive when the main part of the wave is   
outside the corotation radius.
By combination of the sign of the work done on waves and the sigh of
the wave energy, excitation or damping of the waves is determined.

Whether resonance appears in the propagation region of oscillations or not
is also important for observability of the oscillations.
The propagation region of inertial-acoustic oscillations is specified by
$(\omega-m\Omega)^2>\kappa^2$, while that of the g-mode oscillations
is $(\omega-m\Omega)^2<\kappa^2$.
 
In this paper, some of the above results [i.e., the sign of work done on
oscillations by resonance] are derived under the approximation 
that the oscillations are local in the sense that their radial wavelength is
shorter than the characteristic radial scale of disks (see section 5).
The results, however, are free from the approximation, and valid even
in the case of $kr<1$.

The results mentioned so far restrict frequency ranges of resonantly
excited oscillations, but their frequencies are not determined discretely
or uniquely.
Further restrictions concerning excited oscillations are made in section 6
by introducing assumptions that the oscillations are local and thus they are 
predominantly at the boundary of their propagation region, i.e.,
$(\omega-m\Omega)^2\sim \kappa^2$.
Results based on this assumption are summarized in tables 1 and 2 for two 
cases where the disk
deformation is a warp (table 1) and it is a one-armed one plane-symmetric
with respect to the disk plane (table 2).
As shown in these tables, inertial-acoustic and g-mode oscillations are
excited at the radius of $4r_{\rm g}$.
Applications of these results to observed high-frequency QPOs are made, for
example, by Kato and Fukue (2006), Kato (2007) and papers referred therein.

As mentioned before, unless the local approximation of oscillations is
introduced, resonantly excited oscillations are not determined uniquely.
What is determined is possible frequency range of excited oscillations.
This issue is discussed in section 7 and the results are summarized in 
figures 1, 2 and 3.
Eigenfunctions of global trapped oscillations will determine the most
probable frequencies of resonantly excited oscillations.
This problem, however, is outside the scope of this paper.  

Finally, it is important to note that for the resonance considered in this paper
to occur, the disk must be relativistic, i.e., otherwise, the resonant condition
is not satisfied in the disk.
In other words, for resonance to occur in relativistic disks,
the disk deformation must be one-armed 
(i.e., a warp or one-armed pattern symmetric with respect to the equator).

\bigskip\noindent
{\bf Appendix A General Form of Nonlinear Coupling Terms}

In general cases of $\Gamma_1\not= 1$, the pressure coupling terms,
$\mbox{\boldmath $C$}_{\rm p}$, given by equation (\ref{2.6})
should be modified, although the term of 
$\mbox{\boldmath $C$}_{\psi}$ is unchanged.
From equations (9)--(14) of Kato (2004), we have, after some manipulations,
\begin{eqnarray}
     \rho_0C_{p,k}({\mbox{\boldmath $\xi$}}, {\mbox{\boldmath $\xi$}})
     =&&-{\partial\over\partial r_i}\biggr(p_0{\partial\xi_i\over\partial r_j}
                                                             {\partial\xi_i\over\partial r_i}\biggr)
       +{\partial\over\partial r_j}\biggr[(\Gamma_1-1)p_0
                          {\partial\xi_j\over\partial r_k}
                          {\partial \xi_i\over\partial r_i}\biggr]     \nonumber\\
        +&& {1\over 2}{\partial\over\partial r_k}\biggr[(\Gamma_1-1)p_0
                          {\partial\xi_i\over\partial r_j}
                          {\partial \xi_j\over\partial r_i}\biggr]     
       +{1\over 2}{\partial\over\partial r_k}\biggr[\Gamma_1(\Gamma_1-1)p_0
                          {\partial\xi_i\over\partial r_i}
                          {\partial \xi_j\over\partial r_j}\biggr].      
\end{eqnarray}
Then, among $\mbox{\boldmath $\xi$}^{(1)}$,
$\mbox{\boldmath $\xi$}^{(2)}$, and $\mbox{\boldmath $\xi$}^{(3)}$,
such commnesurable relations as 
\begin{eqnarray}
     \int\rho_0 \mbox{\boldmath $\xi$}^{(3)}
          \mbox{\boldmath $C$}_{\rm p}
                 ( \mbox{\boldmath $\xi$}^{(1)}, \mbox{\boldmath $\xi$}^{(2)})dV
      && =\int\rho_0 \mbox{\boldmath $\xi$}^{(1)}
          \mbox{\boldmath $C$}_{\rm p}
                 ( \mbox{\boldmath $\xi$}^{(2)}, \mbox{\boldmath $\xi$}^{(3)})dV
      =\int\rho_0 \mbox{\boldmath $\xi$}^{(2)} 
          \mbox{\boldmath $C$}_{\rm p}
                 ( \mbox{\boldmath $\xi$}^{(1)}, \mbox{\boldmath $\xi$}^{(3)})dV
                                    \nonumber \\
       &&= \cdot\cdot\cdot.   
\end{eqnarray}
hold, where $\mbox{\boldmath $\xi$}^{(1)}$, $\mbox{\boldmath $\xi$}^{(2)}$,
and $\mbox{\boldmath $\xi$}^{(3)}$ are arbitrary functions of 
$\mbox{\boldmath $r$}$.
This can be shown by performing the integrations by parts, so that the 
integrands become symmetric with respect to 
$\mbox{\boldmath $\xi$}^{(1)}$, $\mbox{\boldmath $\xi$}^{(2)}$,
and $\mbox{\boldmath $\xi$}^{(3)}$.
This commensurable relation is a general nature of conservative systems.
In the cases where varous perturbations are on a disk, energy exchange among
these perturbations may occur through resonant processes.
The above commensurable relation, however, show that even in such cases, the total 
energy of perturbations is conserved in a conservative system.
In other words, there is no spontaneous growth of perturbations in equilibrium disks.
This is a reason why we consider deformed disks as the case of excitation of oscillations.


\end{document}